\documentclass[preprint,12pt]{elsarticle}
\usepackage{graphics}
\usepackage{graphicx}
\usepackage{subfigure}

\usepackage{latexsym}
\usepackage{amsmath}
\usepackage{graphicx}
\usepackage{times}
\usepackage{graphicx,multicol,enumerate,subfigure,amsmath}
\usepackage{apropos}

\usepackage{fullpage}
\usepackage{setspace}
\usepackage[english]{babel}
\usepackage[version=3]{mhchem}
\usepackage{amsmath, amssymb}
\usepackage{verbatim, latexsym}
\usepackage{colortbl}

\usepackage{soul}
\setstcolor{red}

\biboptions{sort&compress}
\journal{Computers \& Fluids}

\onehalfspace

\begin{document}

\begin{frontmatter}

\title{On the Shape Optimization of Flapping Wings and their Performance Analysis}
\author{\textbf{Mehdi Ghommem}$^{\mbox{1}}$, \textbf{Nathan Collier}$^{\mbox{2}}$, \textbf{Antti H. Niemi}$^{\mbox{3}}$, and \textbf{Victor M. Calo}$^{\mbox{1,2}}$}

\address{$^{\mbox{1}}$ Division of Physical Sciences \& Engineering (PSE) \\
King Abdullah University of Science and Technology (KAUST) \\
Thuwal 23955-6900, Kingdom of Saudi Arabia}

\address{$^{\mbox{2}}$ Center for Numerical Porous Media (NumPor) \\
Division of Computer, Electrical, Mathematical Sciences \& Engineering (CEMSE) \\
King Abdullah University of Science and Technology (KAUST) \\
Thuwal 23955-6900, Kingdom of Saudi Arabia}

\address{$^{\mbox{3}}$ Department of Civil and Structural Engineering \\
Aalto University \\
Espoo FI-02150, Finland}



\begin{abstract}
The present work is concerned with the shape optimization of flapping wings in forward flight. The analysis is performed by combining a gradient-based
optimizer with the unsteady vortex lattice method (UVLM). We describe the UVLM implementation and provide insights on how to select properly the mesh and time-step sizes to achieve invariant UVLM simulation results under further mesh refinement. Our objective is to identify a set of optimized shapes that maximize the propulsive efficiency, defined as the ratio of the propulsive power over the aerodynamic power, under lift, thrust, and area constraints. Several parameters affecting flight performance are investigated and their impact is described. These include the wing's aspect ratio, camber line, and curvature of the leading and trailing edges. This study provides guidance for shape design of engineered flying systems.
\end{abstract}

\end{frontmatter}

\section{Introduction}
The design and optimization of micro-air vehicles (MAVs) have been recently the topic of many research efforts \cite{Stanford2010,Moore,Kurdi2010,Stanford2012,Lian2004,Ghommem2012,GhommemCCP2012,Ghommem2010,Stanford2009,Khan2007,Tuncer2005,Shyy1999,Rakotomamonjy,
Mueller2001,Chimakurthi,Pines2006}. The MAV concept is quite new and the associated design space is large and not fully explored yet. These small flying aeroelastic systems are expected to operate in urban environments and confined spaces (i.e., inside buildings, caves, tunnels) and endure a variety of missions such as inspection and observation of harsh environments inaccessible to other types of vehicles or where there is danger to human life. Consequently, their design must satisfy stringent performance requirements, such as high maneuverability at low speeds, hovering capabilities, high lift to sustain flight, and structural strength to survive gust loads and possible impacts. These requirements can be achieved mainly through two propulsion mechanisms:
rotating helicopter blades or flapping wings \cite{Petricca2011}. Through observing and simulating insects
and birds flights, it has been concluded that flapping wings offer greater efficiency, especially at
small scales~\cite{Shyy2010,Ansari2006,Shyybook}. As a step toward designing efficient flapping-wing vehicles, several studies~\cite{Willmott1997,Tobalske2003,Wallace2006,Sane2003,Dickinson1999,Muijres2008,Tobalske2007,Tian2006,GhommemCCP2012} explored and investigated the characteristics of these flying animals.

Although it is well known that natural flyers exploit a variety of mechanisms and aerodynamic aspects to control and manoeuver their flights, experimental observations do not enable a complete understanding of the physical aspects and dynamics of flapping flight. As such, there is a need to model the unsteady aerodynamic aspects of flapping-wing vehicles. Computational modeling and simulation are necessary to evaluate
the performance requirements associated with flapping flight and to identify the relative impact of different design parameters (e.g., flapping parameters, shape characteristics). Several computational modeling strategies that are based on variable fidelity physics have been reported in the literature \cite{Stanford2010,Bansmer2012,Persson2012,Ghommem2010,Berman2007,Yuan2010,Liu2009,Sane2002,Willis2007}. Willis et al. \cite{Willis2008} developed a multifidelity computational framework that involves a combination of potential flow models and a Navier-Stokes solver. They showed how the use of different levels of geometric and physical modeling fidelity can be exploited to ease the design process of flapping wing systems. Certainly, the higher-fidelity Navier-Stokes simulations incorporate a more complete physical model of the flapping flight problem, however, the extensive computational resources and time associated with the use of these tools limit the ability to perform optimization and sensitivity analyses in the early stages of MAV design. Thus, to alleviate this burden and enable rapid and reasonably accurate exploration of a large design space, it is fairly common to rely upon a moderate level of modeling fidelity to traverse the design space in an economical manner \cite{Stanford2011,Stanford2012}. As such, several research efforts have considered the use of the unsteady vortex lattice method (UVLM) for the design of avian-like flapping wings in forward flight \cite{VestKatz,Fritz2004,Stanford2010,Ghommem2012,Smith1996,GhommemCCP2012,Persson2012}.

The present work is concerned with shape optimization of flapping wings in forward flight. Our approach to the problem combines a local gradient-based
optimizer with UVLM. The optimizer is based on the globally convergent (to a stationary point, not necessarily a global solution) method of moving asymptotes (GCMMA) \cite{Svanberg1987,Svanberg2002}. It belongs to the class of sequential approximate optimization methods and employs conservative convex separable approximations for solving inequality constrained nonlinear programming problems. It searches by generating approximate subproblems at each iteration, in which both the objective and constraint functions are replaced by convex functions. The construction of these approximating functions is based mainly on gradient information at the current iteration point in the design space.

 The unsteady vortex lattice method computes the forces generated by pressure differences across the wing surface resulting from acceleration- and circulation-based phenomena. This accounts for unsteady effects such as added mass forces, the growth of bound circulation, and the wake. Since the formulation of UVLM requires that fluid leaves the wing smoothly at the trailing edge (through imposing the Kutta condition), it does not cover the cases of flow separation at the leading-edge and extreme situations where strong wing-wake interactions take place. In other words, UVLM applies only to ideal fluids, incompressible, inviscid, and irrotational flows where the separation lines are known a priori. Nevertheless, Persson et al. \cite{Persson2012} showed through a detailed comparison between UVLM and higher-fidelity computational fluid dynamics simulations for flapping flight that the UVLM schemes produce accurate results for attached flow cases and even remain trend-relevant in the presence of flow separation. As such, in \cite{Persson2012} the authors recommended the use of an aerodynamic model based on UVLM to perform preliminary standard design studies of flapping wing vehicles, specifically, for cruise configurations where there is no flow separation nor substantial wing-wake interactions that would degrade the performance of the vehicle. The reduced fidelity afforded by UVLM makes its associated simulation time for each flow configuration is on the order of minutes on a desktop.

In this paper, we seek to identify a set of optimized shapes that maximize the propulsive efficiency of a flapping wing under several constraints. We explicitly include lift, thrust, and area constraints. The wing geometry is described using B-spline parameterizations, which are the standard technology for describing geometries in computer-aided design (CAD)~\cite{Rogers01,PiegTil97,Farin95,CoRiEl01}. This choice simplifies the design optimization process by enabling mesh generation directly from the CAD model. The associated basis functions can be used to smoothly discretize wing shapes with few degrees of freedom. The location of the control points constitutes the design variables. Results suggest that changing the shape yields significant improvement in the flapping wings performance. This study is our first step towards constructing a unified framework that will facilitate the design of engineered flying systems.

The remainder of the paper is organized as follows: first, the aerodynamic model used to simulate flapping flights is detailed. The B-spline representation that defines the wing geometry is then introduced, followed by the formulation of the shape optimization problem. The optimal shapes obtained for different flight configurations are reported and insights of the trailing and leading edges curvatures, camber, and aspect ratio effects on the flapping flight performance are presented. The work concludes with a discussion of the main features of the optimal shapes.

\section{Aerodynamic Modeling of Flapping Wings}
We use a three-dimensional version of the unsteady vortex lattice method (3D UVLM) to simulate the
aerodynamic response of flapping wings in forward flight. This aerodynamic tool is capable of simulating incompressible and inviscid flow past moving thin wings and capturing the unsteady effects of the wake, but not the viscous effects, flow separation at the leading-edge, and extreme situations with strong wing-wake interactions. Unlike standard computational fluid dynamics schemes, this method requires meshing of the wing planform only and not of the whole flow domain. UVLM uses a global vorticity circulation balance, which is described on a vortex lattice. Given the circulation on a ring, UVLM relies on the Biot-Savart law to construct the velocity field at every point in 3D space. Using the wing surface discretization, UVLM imposes a collocated no-penetration condition to construct a potential flow about the mid-surface of the wing. The satisfaction of the no-penetration condition allows us to compute the circulation in each vortex loop, closing the system. By this sequence of approximations, UVLM reduces the demand for computational resources. These features make the method competitive for performing optimization studies that require many simulations by appropriately capturing flow dynamics at a low computational cost.

Features of the UVLM solver used in this study include
\begin{itemize}
  \item The shape of the wing is generated using a B-spline representation.
  \item The wing surface is discretized into a lattice of vortex rings. Each vortex ring consists of four short straight vortex segments, with a collocation point placed at its center.
  \item For rigid wings, the position of the grid points is specified by applying a sequence of rotations (pitching and flapping).
  \item The no-penetration condition is imposed at collocation points. The normal component of the velocity due to wing–-wing interactions, wake-–wing interactions, free-stream velocities, and wing rotations is assumed to vanish at each collocation point. Using the Biot-Savart law to compute velocities in terms vorticity circulations $\boldsymbol{\Gamma}$ yields a linear system of equations that can be expressed as:
  \begin{eqnarray}\label{NPC}
  \bA^{\mathrm{wi-wi}}\boldsymbol{\Gamma}_\mathrm{wi} = - \bA^{\mathrm{wa-wi}}\boldsymbol{\Gamma}_\mathrm{wa} + \bV_\mathrm{n}
  \end{eqnarray}
where $\bA^\mathrm{wi-wi}$ and $\bA^\mathrm{wa-wi}$ are wing-wing and wake-wing influence matrices, respectively. The vector $\bV_\mathrm{n}$ collects the normal component of the velocity at each collocation point due to the wing motion. The vectors $\boldsymbol{\Gamma}_\mathrm{wi}$ and $\boldsymbol{\Gamma}_\mathrm{wa}$ stand for the circulations of the vortex elements on the wing and wake, respectively~\cite{Ghommem2012,Stanford2010}. The LAPACK library \cite{Lapack} is used to solve the linear system given by Equation \eqref{NPC}.
  \item Vorticity is introduced to the wake by shedding vortex segments from the trailing
edge. These vortices are moved with the fluid particle velocity
and their individual circulation remains constant (i.e., $\boldsymbol{\Gamma}_\mathrm{wa}^{t+\Delta t}=\boldsymbol{\Gamma}_\mathrm{wa}^{t}$). The wake elements have been truncated in the flowfield and load computation and only those which are located within 10 $c$ are accounted for, where $c$ is the chord length. This truncation was observed to speed up significantly the simulation while introducing negligible loss in the solution accuracy.
\item The pressure is evaluated at each collocation point based on the unsteady Bernoulli equation and then integrated over the wing surface to compute the aerodynamic forces and power.
\end{itemize}

Further details of the derivation, implementation, and verification of this method and aerodynamic loads computation are
provided in~\cite{KatzPlotkin,Stanford2010,Ghommem2012,Ghommem2012B,Persson2012,Willis2007}.

\section{Modeling of Wing Shape: B-spline Representation}
\subsection{B-spline curves and surfaces}
B-splines are piecewise polynomial functions generalizing the Bernstein-B\'{e}zier polynomials. The B-spline basis functions of degree $p$, denoted by $N_{i,p}(\xi)$, associated
to a non-decreasing set of coordinates, called the knot vector $\boldsymbol{\cX} =
\{\xi_1,\xi_2,\ldots,\xi_{n+p+1}\}$, are defined recursively as
\[
\begin{aligned}
N_{i,0}(\xi) &= \begin{cases}
1, & \xi_i \leq \xi < \xi_{i+1} \\
0, &\text{otherwise}
\end{cases} \\
N_{i,p}(\xi) &= \frac{\xi-\xi_i}{\xi_{i+p}-\xi_i}N_{i,p-1}(\xi) + \frac{\xi_{i+p+1}-\xi}{\xi_{i+p+1}-\xi_{i+1}}N_{i+1,p-1}(\xi), \quad p>0
\end{aligned}
\]
for $i=1,\ldots,n$ and $p \geq 1$.  Knot multiplicities reduce the continuity of the basis at the
location of the multiplicity. If a knot multiplicity of $k$ is used,
the continuity on the basis is $C^{p-k}$ at that knot. The basis becomes interpolatory at knots with multiplicity $p$ whereas
knot multiplicity of $p+1$ makes the basis discontinuous.

The B-spline curve of degree $p$ with control points
$\bP_1,\ldots,\bP_n$ is defined on the interval $[a,b] =
[\xi_{p+1},\xi_{n+1}]$ as the linear combination of the control points
and basis functions, that is,
\[
\bC(\xi) = \sum_{i=1}^n N_{i,p}(\xi) \bP_i.
\]
The piecewise linear interpolation of the control points is called the
control polygon. A feature of B-spline curves is that the curve
defined by the basis and control points lies inside the convex
hull of the control polygon. This makes the control polygon useful for
approximating the rough character of a curve.

A B-spline surface is defined using tensor products of B-spline basis
functions written in two parametric coordinates $\xi$ and $\eta$. Let
$N_{i,p}$ and $M_{j,q}$ denote basis functions of degree $p$ and $q$
associated to the knot vectors $\boldsymbol{\cX} = \{
\xi_1,\xi_2,\ldots,\xi_{n+p+1} \}$ and $\boldsymbol{\cY} =
\{\eta_1,\eta_2,\ldots,\eta_{m+q+1}\}$, and $\bP_{ij}$, $i=1,\ldots,n$,
$j=1,\ldots,m$ is a net of control points in three-dimensional space. A B-spline surface is then defined as
\[
\bS(\xi,\eta) = \sum_{i=1}^n \sum_{j=1}^m N_{i,p}(\xi) M_{j,q}(\eta) \bP_{ij}.
\]
B-splines have long been used in the computer aided design community
to model curves and surfaces. The reader is referred to~\cite{Rogers01,PiegTil97,Farin95,CoRiEl01} for more details on
B-splines.

\subsection{Wing shape parametrization}
\begin{figure}[ht]
  \begin{center}
      \subfigure[Undeformed shape]{\includegraphics[width=0.48\textwidth]{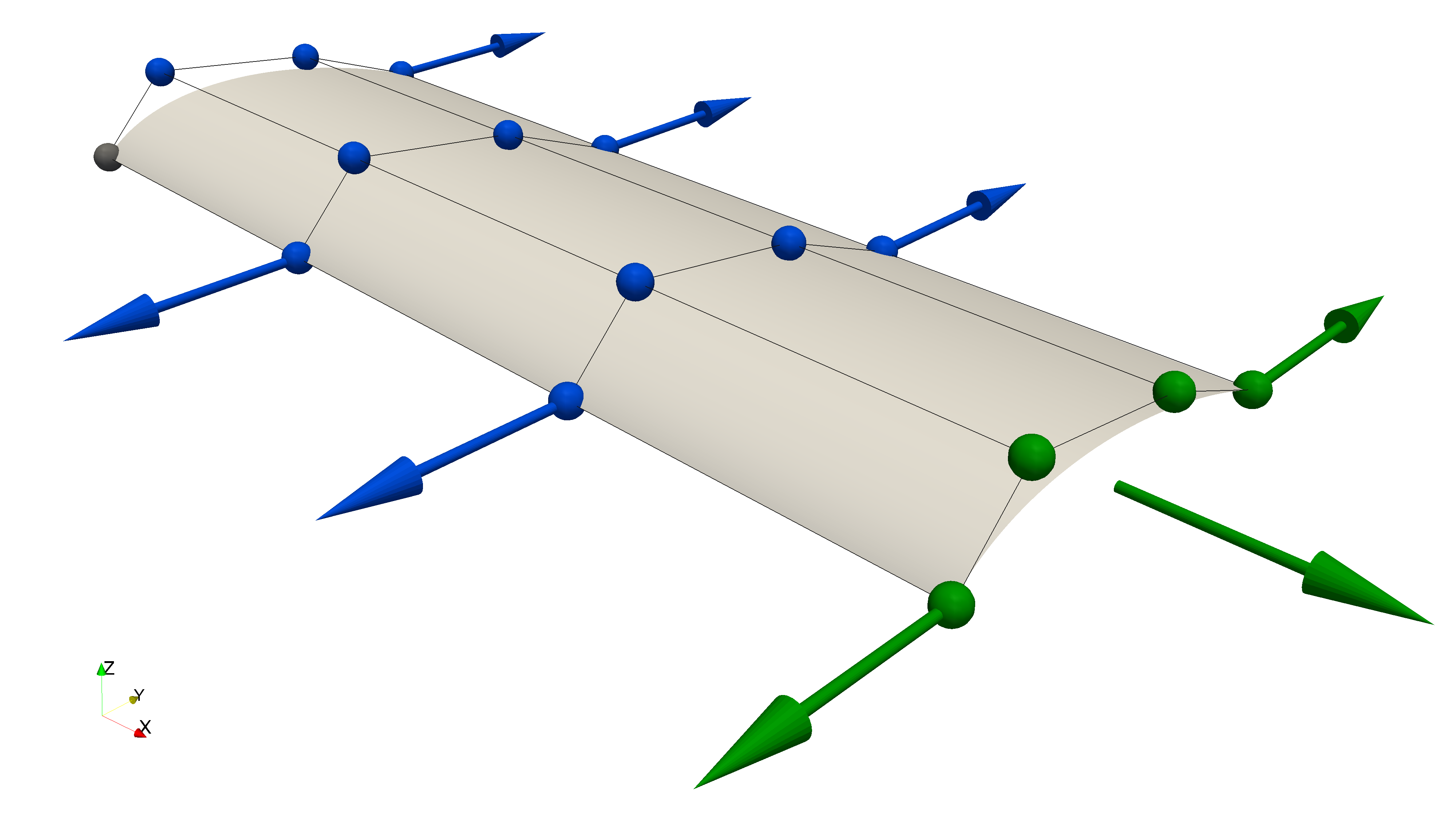}\label{Skematic}}
      \subfigure[Deformed shape]{\includegraphics[width=0.48\textwidth]{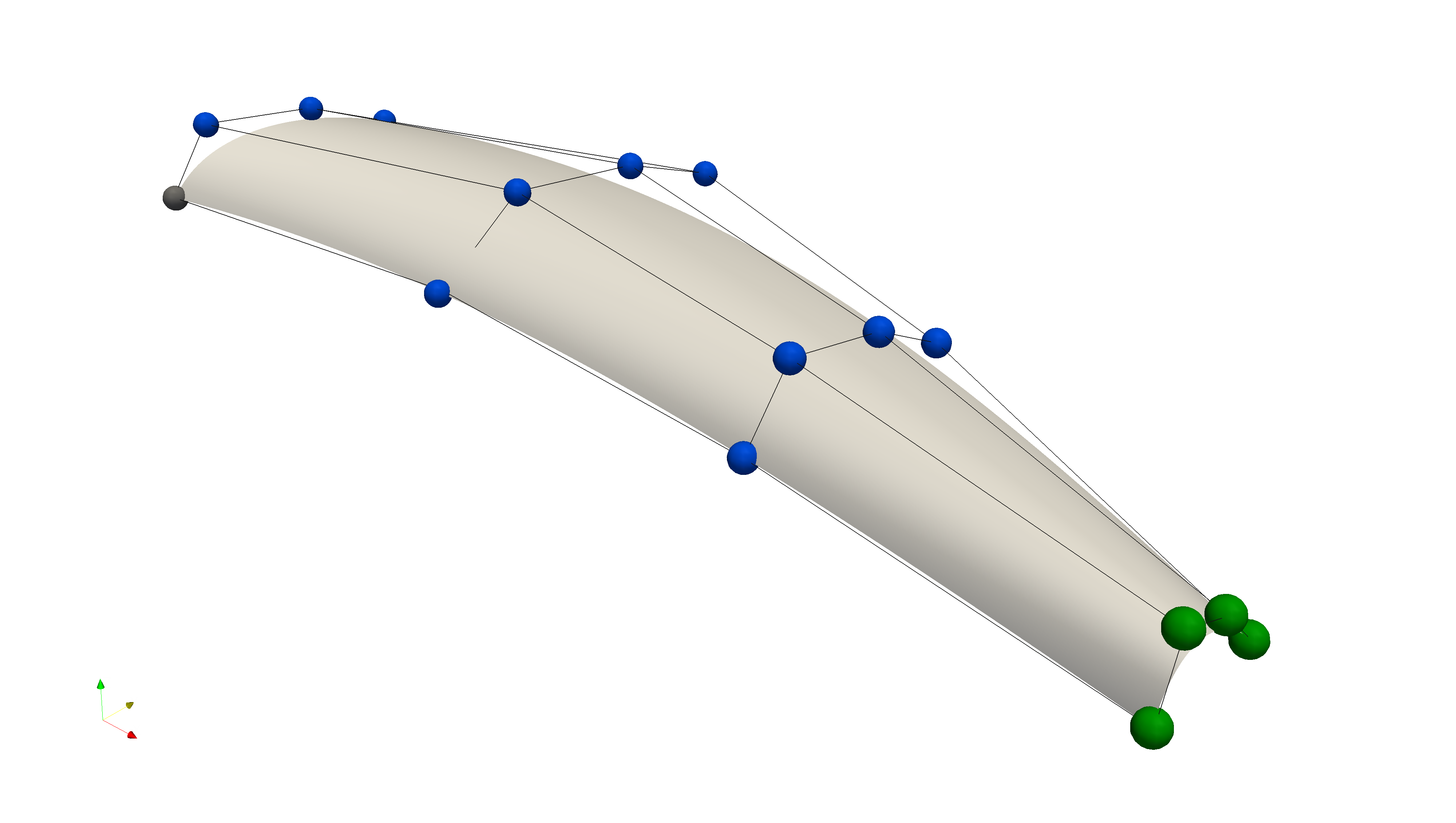}\label{Skematic1}}
      \subfigure[Undeformed shape - top view]{\includegraphics[width=0.48\textwidth]{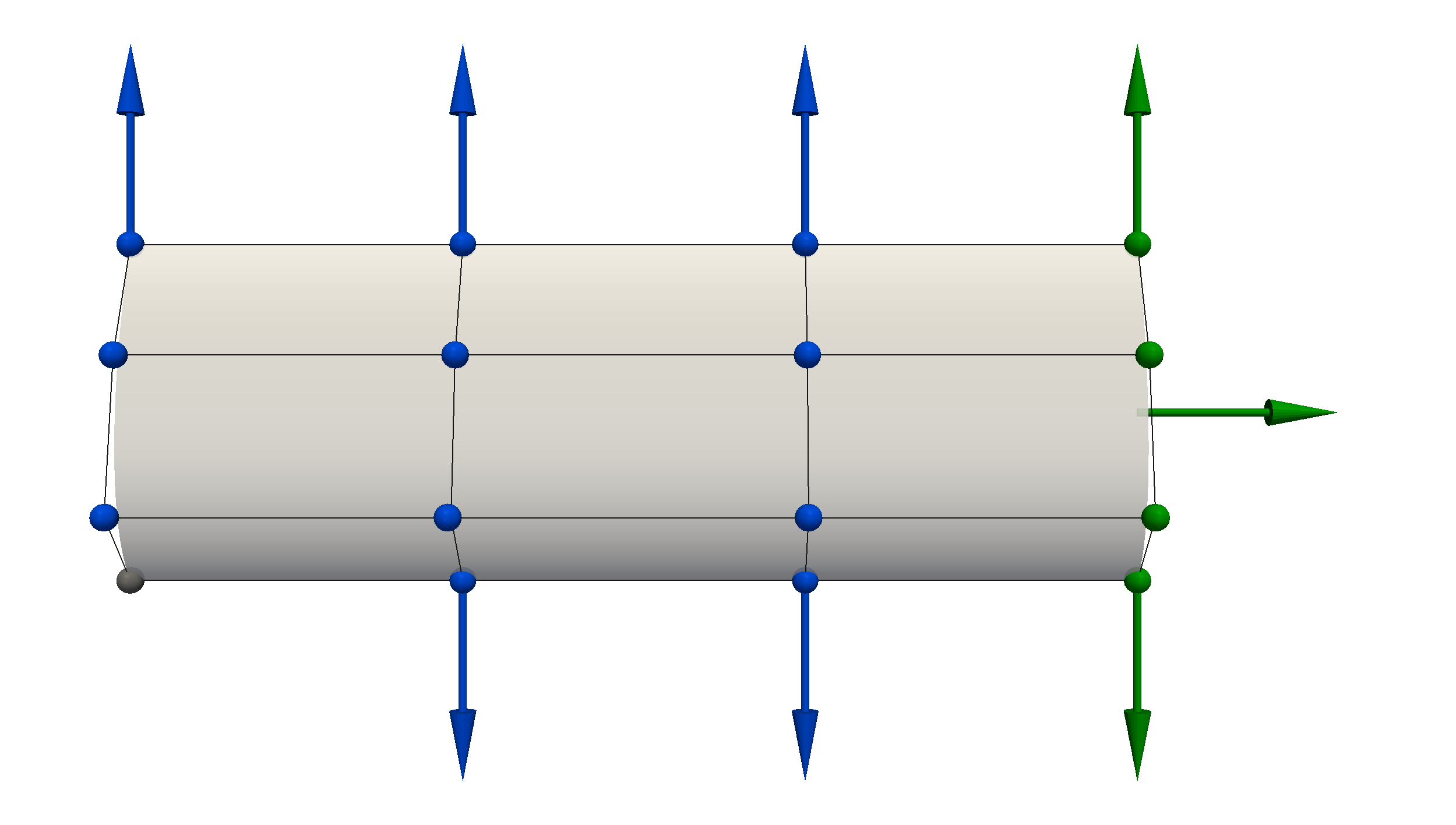}\label{Skematicm}}
      \subfigure[Deformed shape - top view]{\includegraphics[width=0.48\textwidth]{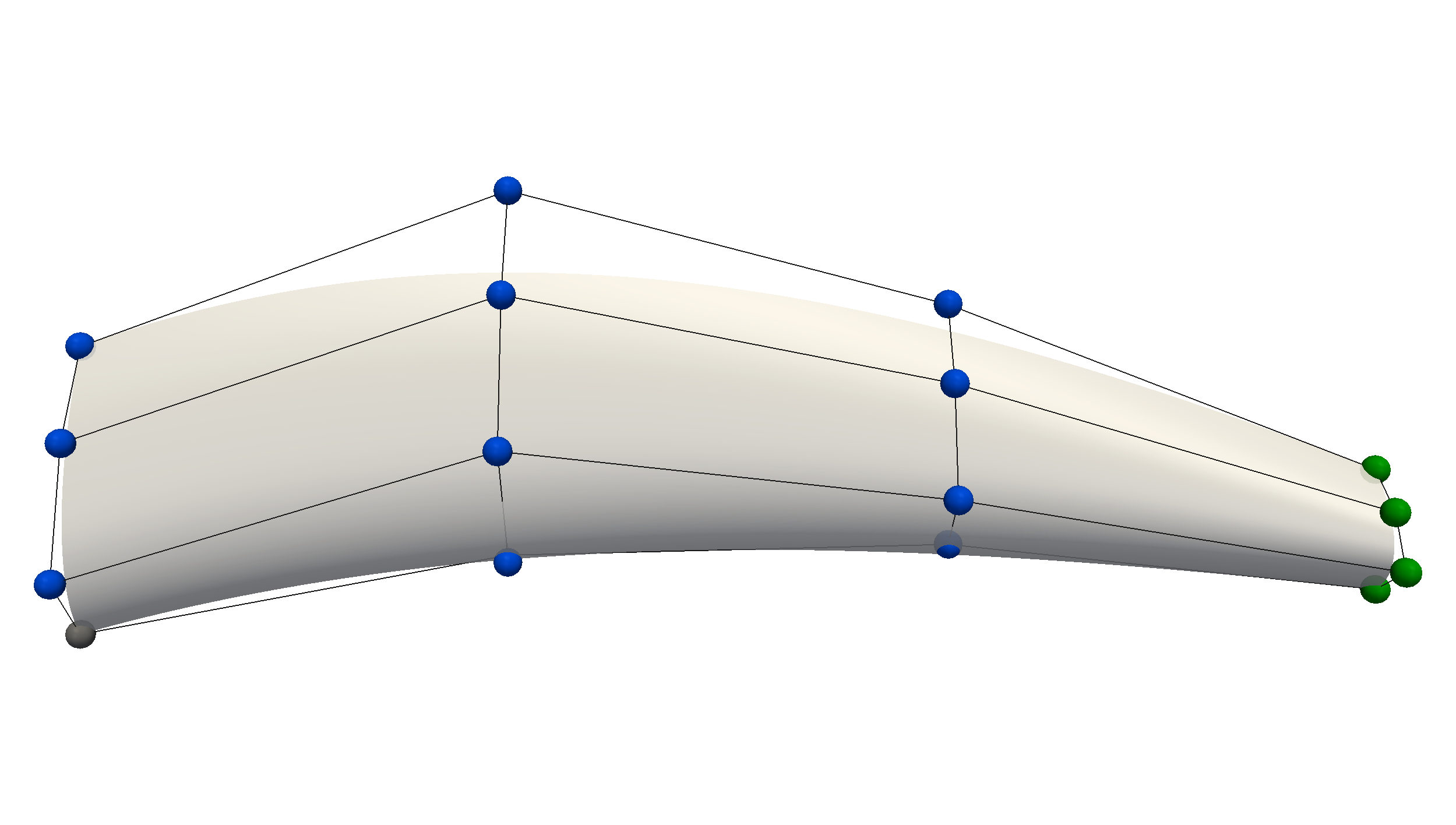}\label{Skematicm1}}
  \end{center}
  \caption{Wing geometry model based on B-spline representation. Spheres are used to represent the control points and arrows denote their perturbation directions. The straight line going over the spheres on the grid is the control polygon while the shaded surface represent the actual wing surface.}
  \label{Skematic}
\end{figure}

In Figure~\ref{Skematic}, we plot the wing geometry model based on a
B-spline representation. The control points are shown as a network of
connected spheres and the wing as a shaded surface. In B-splines, the
surface does not interpolate the control points, rather the
control polygon is an approximation to the surface which it
discretizes. Perturbing the control points changes the shape of
the wing. However, we deform the wing shape in such a
way that it preserves some of its original features. In particular, we
allow the camber edge to scale and translate in the $y$-direction
only, as indicated by the blue spheres and arrows in the figure. The
green spheres and arrows are similar except that we also allow
translation in the $x$-direction. The black sphere is held fixed in
all cases. Finally, we proportionally redistribute the nodes to preserve to some extent the aspect ratio of each element 
 of the mesh. We deform the wing surface in this way for three reasons.
\begin{enumerate}
\item The camber edge is typically something known from experience and
  we do not want to depart completely from established
  designs.
\item Changing the camber significantly may lead to cases where,
  physically, flow separation occurs. This is a case we wish to avoid
  in practice as well as in our numerical method.
\item The approach reduces the number of degrees of freedom such that
  the optimization is more computationally efficient. While
  Figure~\ref{Skematic} uses 48 variables (16 control points in 3D
  space), we define our deformed wing using only 8 degrees of freedom
  (far end scales only, the middle rows scale and translate, and the
  near end scales, and translates in two directions).
\end{enumerate}
As an example, Figures~\ref{Skematic1} and \ref{Skematicm1} depict a possible deformed wing
geometry based on our rules. We modify the wing shape in this way in
all simulation results reported in this work unless stated otherwise.

\section{Shape Optimization}
\subsection{Baseline wing shape}
\begin{figure}
  \begin{center}
      \subfigure[Lift coefficient]{\includegraphics[width=0.68\textwidth]{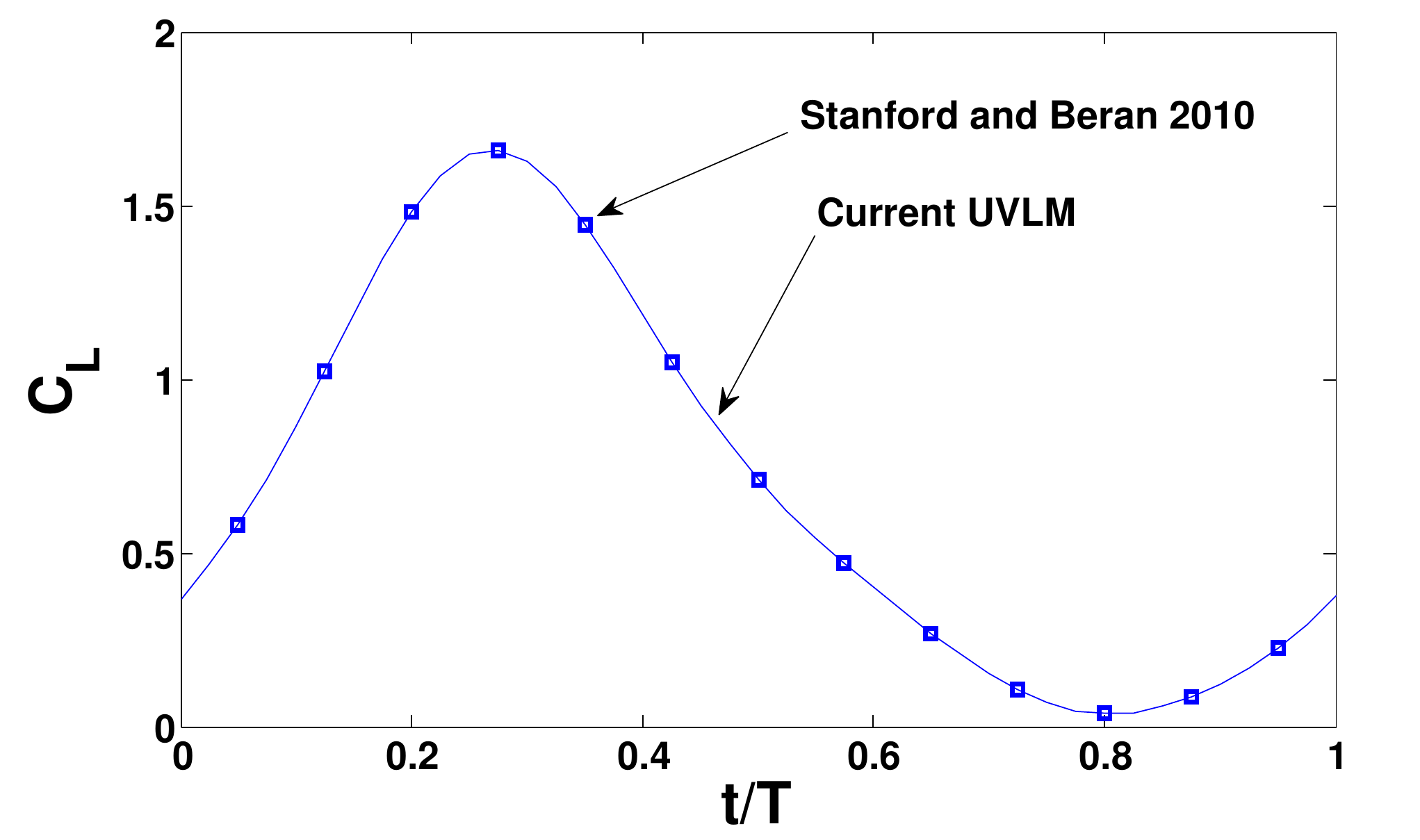}\label{Fig1a}}
      \subfigure[Thrust coefficient]{\includegraphics[width=0.68\textwidth]{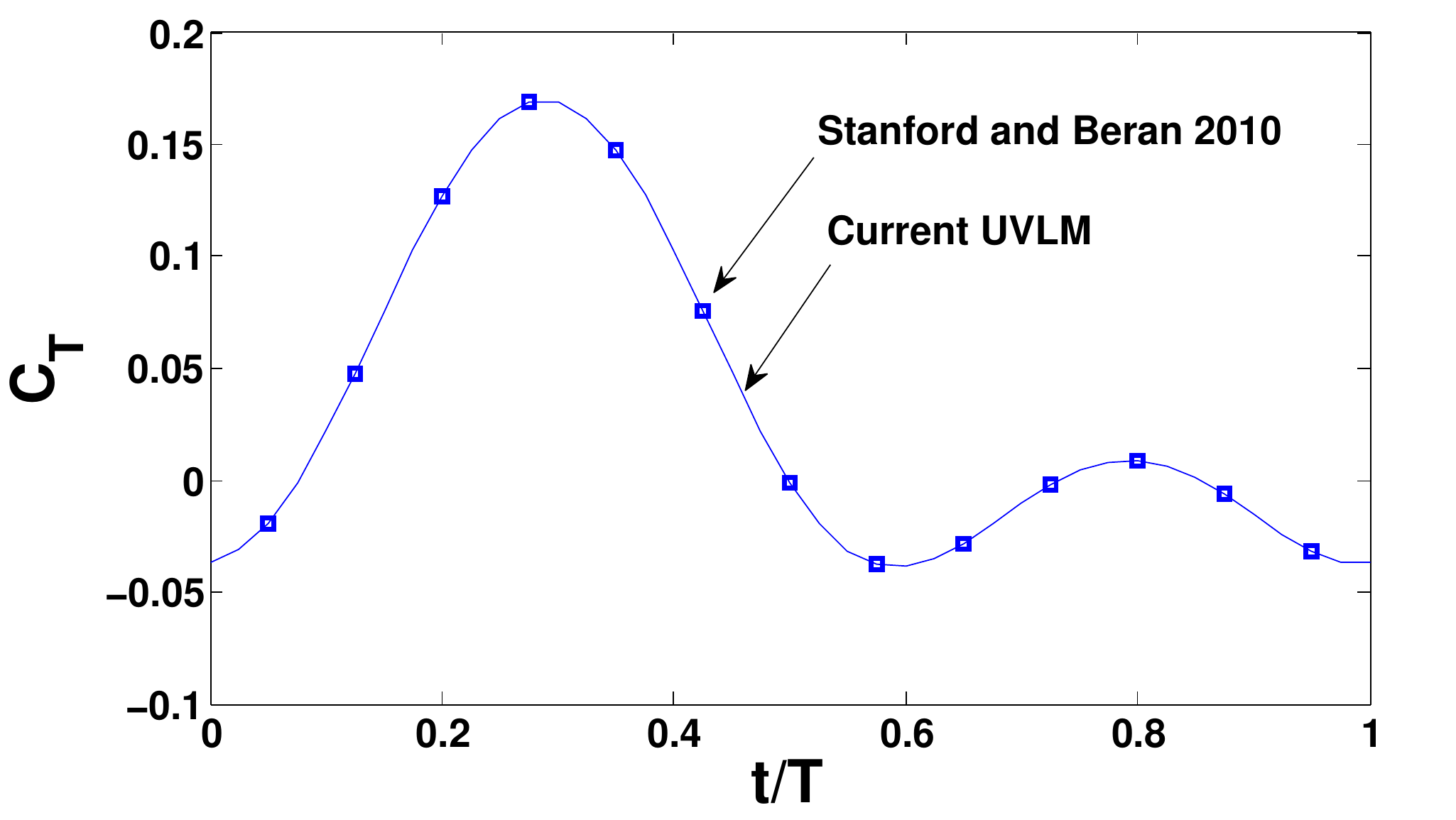}\label{Fig1b}}
  \end{center}
  \caption{Lift and thrust computed from UVLM for one flapping cycle: comparison with results obtained by Stanford and Beran \cite{Stanford2010}. Six elements are used along the chordwise direction and ten are used along the spanwise direction (for half wing).}
  \label{FigComp}
\end{figure}

In this section, we consider cambered rectangular wings with an aspect ratio of six (baseline case). The cambered wing has a NACA 83XX cross-sectional profile as studied previously by Stanford and Beran \cite{Stanford2010} and later by Ghommem et al. \cite{Ghommem2012}. The indication of the last two digits of the NACA profile, describing the wing thickness, is irrelevant for our study since the current implementation of UVLM assumes a thin wing. The symmetric flapping motion (about the wing root) is prescribed as:
\begin{eqnarray}
\phi(t) = A_{\phi} \cos(\omega\;t),
\end{eqnarray}
where $\phi$ is the flapping angle and the flapping amplitude $A_{\phi}$ is set equal to 45$^{\circ}$. Furthermore, the wing root is placed at a fixed angle of attack (pitch) of 5$^{\circ}$. Different reduced frequencies are used to examine the effect of flow unsteadiness on the aerodynamic performance of flapping wings. The reduced frequency is defined as \[\kappa=\frac{\omega}{U_{\infty}} \cdot \frac{c}{2},\] where $\omega$ is the flapping frequency, $c$ is the chord length, and $U_{\infty}$ is the freestream velocity. The transient variations of the lift and thrust over one flapping cycle predicted by the current UVLM and those obtained by Stanford and Beran \cite{Stanford2010} are shown in Figure \ref{FigComp}. A good agreement can be clearly observed. For the sake of comparison, six elements are used along the chordwise direction and 10 are used along the semispan, and the flapping period is discretized into 40 time steps. In the next section, we show the need to consider smaller time steps and more refined aerodynamic meshes to obtain convergence. Our analysis provides guidance to the mesh size and time step size that need to be used to achieve convergent UVLM simulations under refinement. To our knowledge, the analysis presented in the next section is the first that achieves solution independence to discretization for UVLM. The rectangular flapping wing case, as described above, constitutes the baseline case of the optimization studies conducted next.

\subsection{Convergence analysis}
\begin{figure}[ht]
\begin{center}
\includegraphics[width=0.6\textwidth]{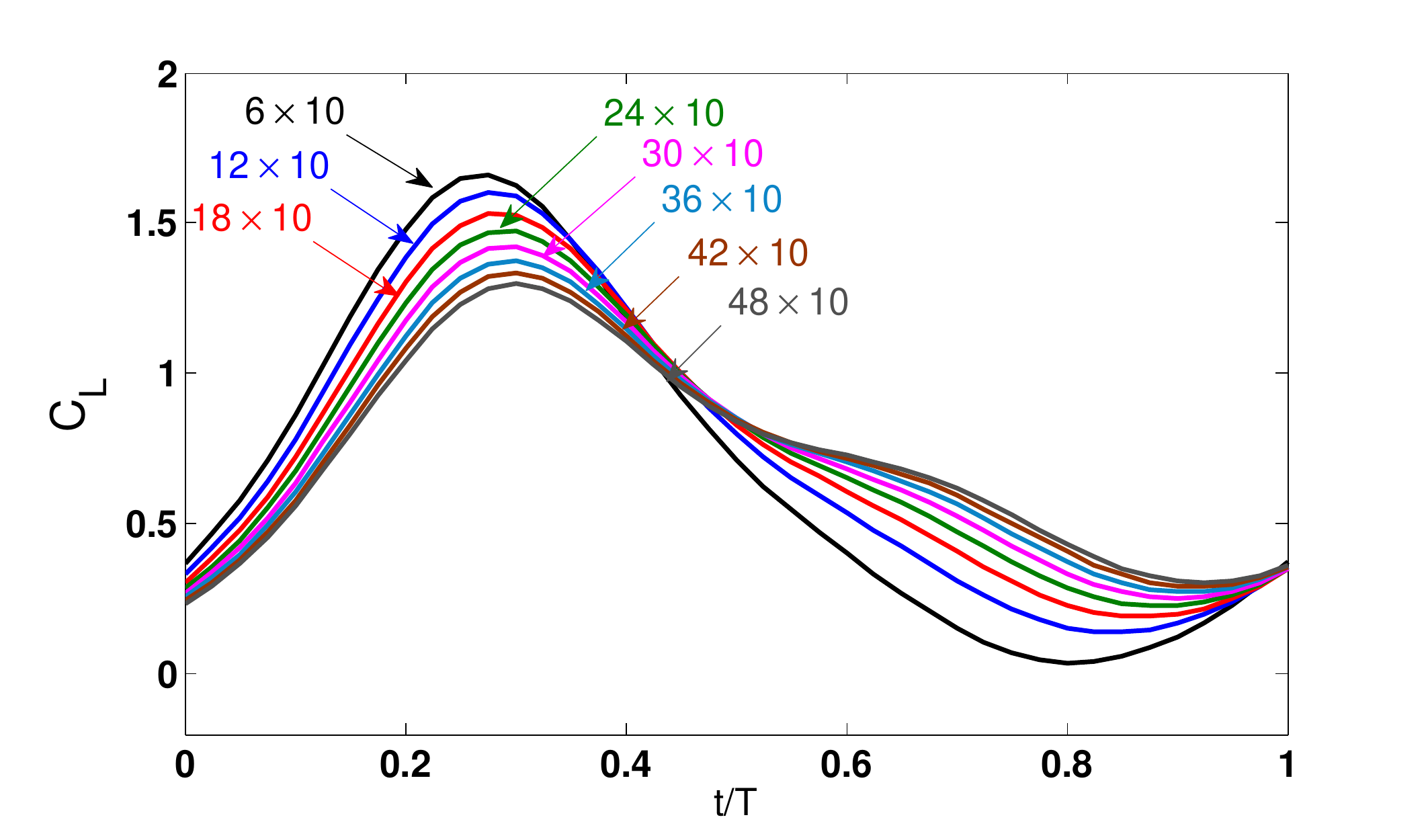}
\end{center}
\caption{Transient variations of lift coefficient for different aerodynamic mesh sizes (fixed number of time steps per flapping cycle, $N_\mathrm{ts}=40$).} \label{Convm1}
\end{figure}

We first investigate the effect of mesh refinement on the lift coefficient while keeping the number of time steps per flapping cycle $N_\mathrm{ts}$ constant and equal to 40. Figure~\ref{Convm1} shows that UVLM fails to converge as the mesh is refined.

\begin{figure}[ht]
\begin{center}
\includegraphics[width=0.6\textwidth]{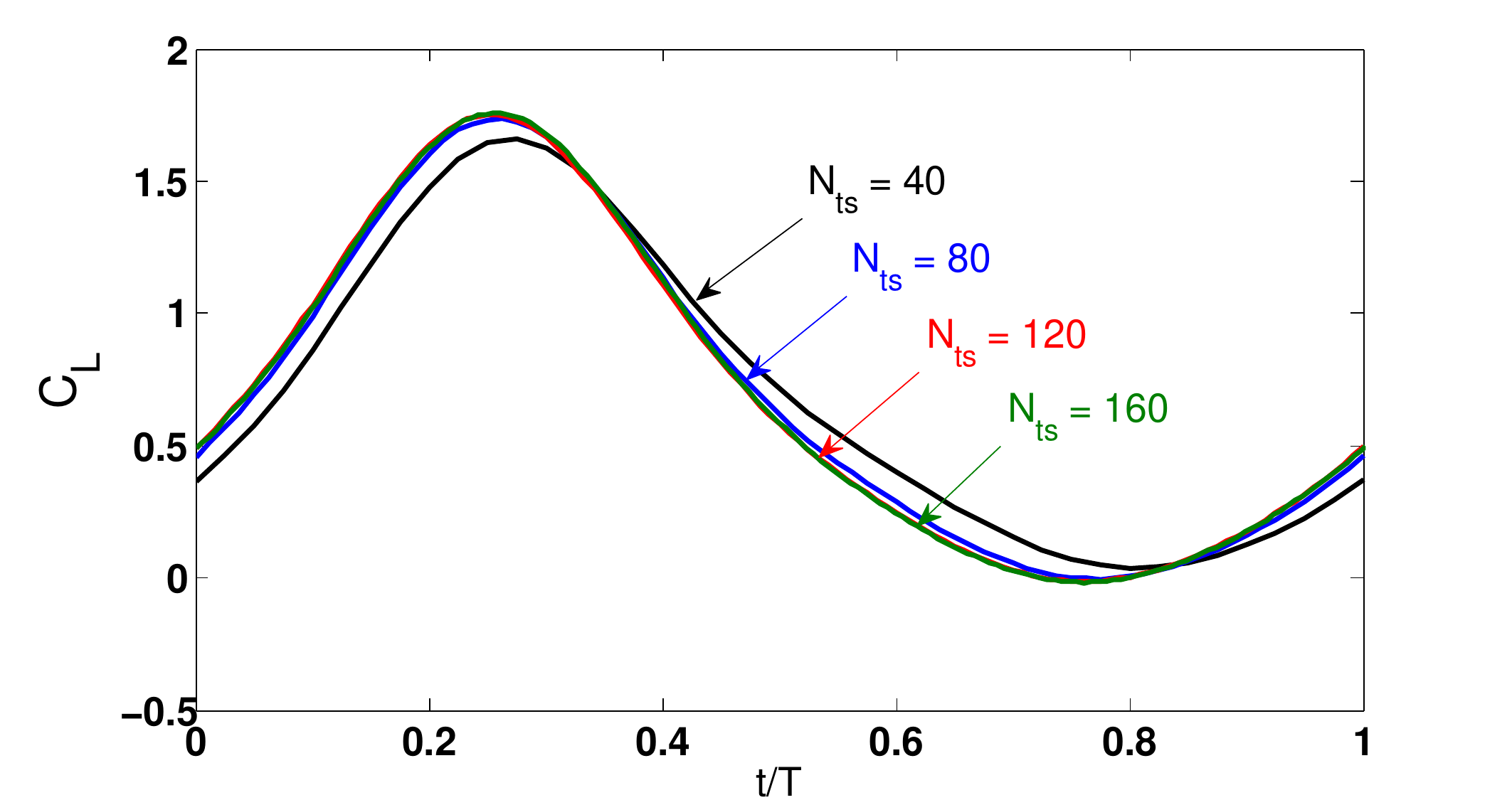}
\end{center}
\caption{Transient variations of lift coefficient for different time step sizes (fixed spatial resolution, 6$\times$10 mesh).} \label{Convt}
\end{figure}

To fix the convergence issue, we perform several simulations while varying the number of time steps per flapping cycle $N_\mathrm{ts}$ over a range of 40 to 160 and compare the transient variations of the lift coefficient to select properly the time step size. The obtained results for a coarse mesh of six chordwise elements and 10 spanwise elements are shown in Figure \ref{Convt}. UVLM converges with a number of time steps per flapping cycle of 120. There is little change between the simulation using a number $N_\mathrm{ts}$ of 120 and that using 160 time steps.

\begin{figure}
  \begin{center}
      \subfigure[Chordwise refinement]{\includegraphics[width=0.48\textwidth]{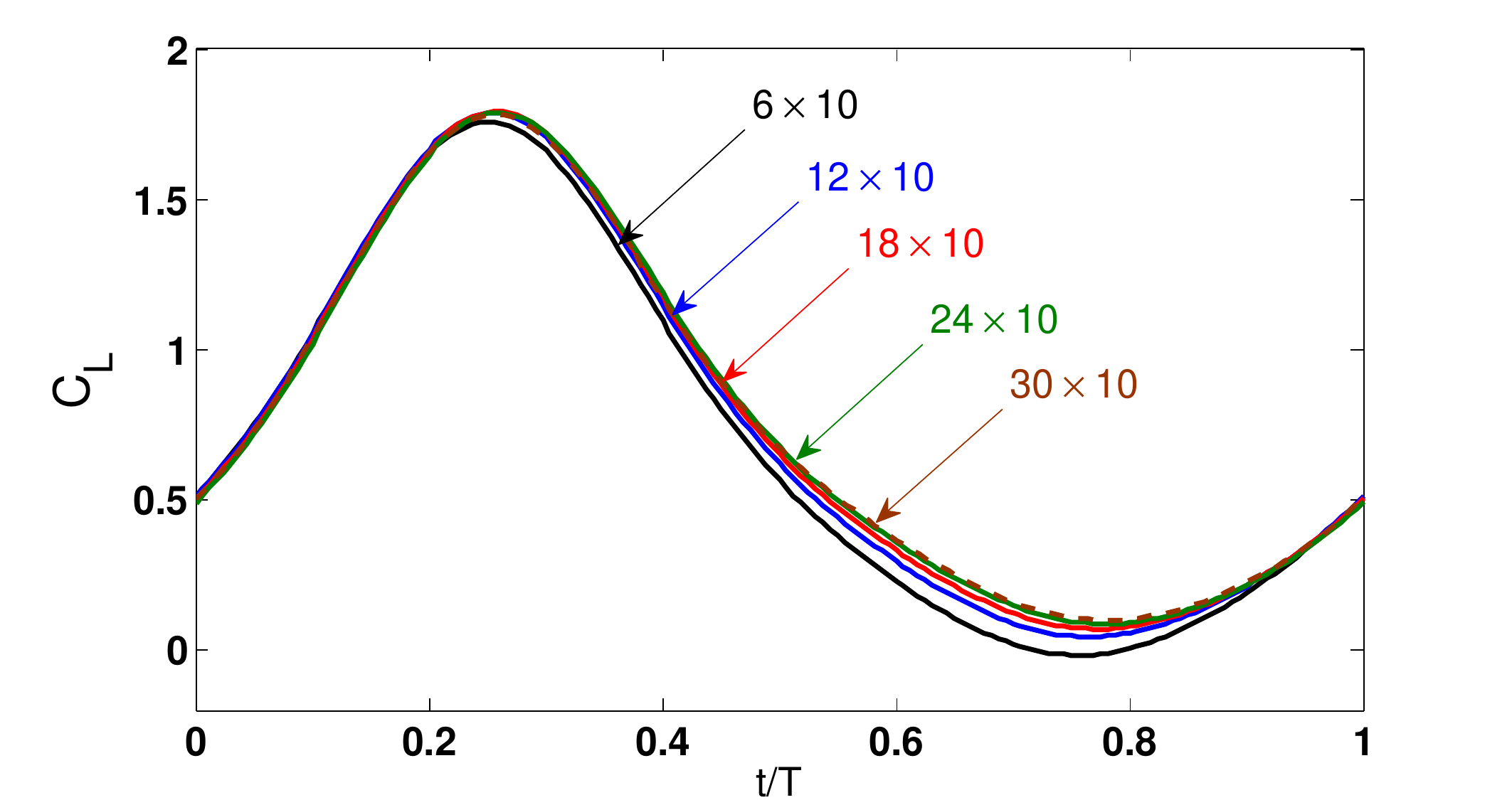}\label{Conv1}}
      \subfigure[Spanwise refinement]{\includegraphics[width=0.48\textwidth]{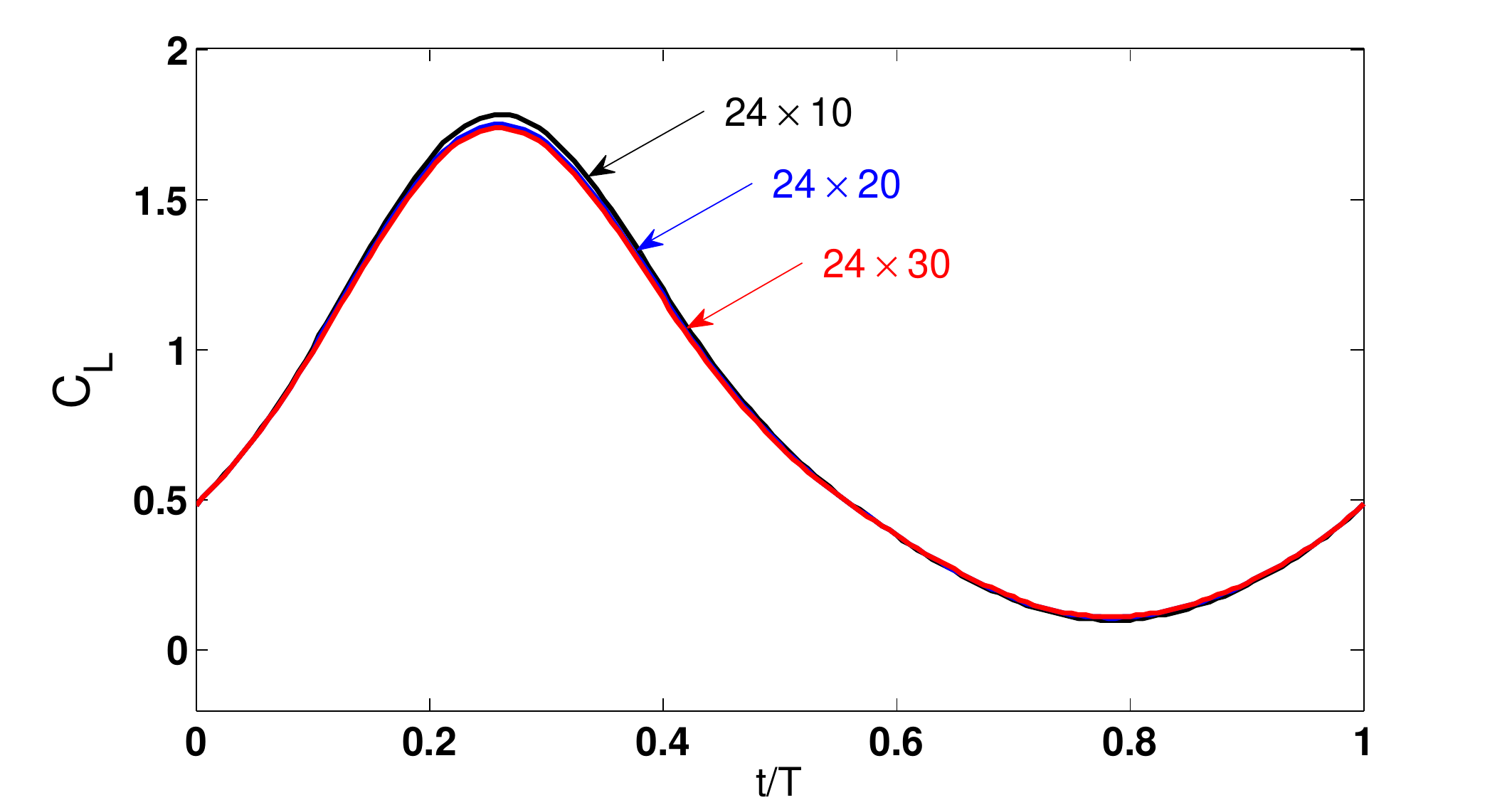}\label{Conv2}}
  \end{center}
  \caption{Transient variations of lift coefficient for different chordwise and spanwise refinements of the aerodynamic mesh.}
  \label{Convm}
\end{figure}

As for the effect of the aerodynamic mesh on the prediction of the aerodynamic loads, the transient variations of the lift coefficient are evaluated for different chordwise and spanwise refinements while keeping $N_\mathrm{ts}$ constant and equal to 120. Results of the lift coefficient obtained for different numbers of chordwise and spanwise elements are shown in Figure~\ref{Convm}. Clearly, refining the mesh along the chordwise direction affects substantially the prediction of the aerodynamic loads. We observe that the use of 24 chordwise panels yields convergence of the lift coefficient. The effect of the spanwise refinement on the convergence trend of the lift force is much less significant, as shown in Figure \ref{Conv2}.

We propose to first determine the appropriate time step in order to observe convergence of UVLM simulations when refining the aerodynamic mesh. For the reminder of this work, the flapping cycle is discretized into 120 time steps and an aerodynamic mesh of 24 chordwise elements and 20 spanwise elements is used. We note that the use of 120 time steps per flapping cycle was checked for the aforementioned refined mesh size and was found appropriate to reach solution convergence; that is, as $N_\mathrm{ts}$ is increased, the simulation results remain unaffected (not shown here).

\subsection{Problem formulation}
\begin{figure}[ht]
\begin{center}
\includegraphics[width=0.65\textwidth]{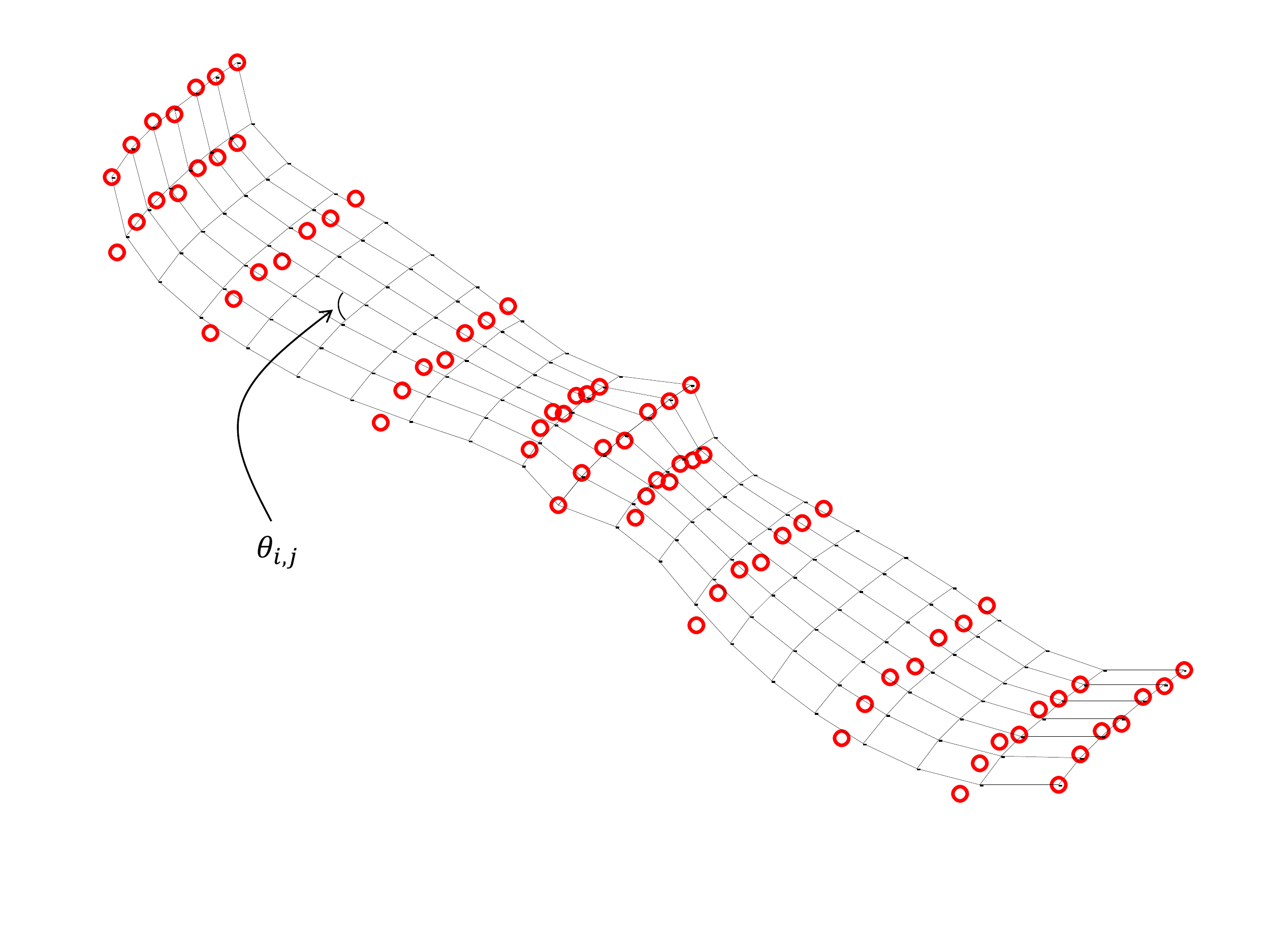}
\end{center}
\caption{Aerodynamic grid of the wing modeled using B-spline. Red circles represent the locations of the control points used to define the wing geometry.} \label{Aerogrid}
\end{figure}

Wing shape selection presents a critical determinant of performance in flapping flight. In this study, we conduct series of shape optimization studies aiming at identifying a suitable set of wing planforms that enables efficient flights. The optimization problem is formulated as follows:
\begin{equation}
\begin{alignedat}{2}
&\text{Maximize} \quad & &\eta(\bx) \\
&\text{Subject to} \quad & &\overline{L^*}(\bx) \geq \overline{L^*}_\mathrm{bl}, \\
&&&\overline{T^*}(\bx) \geq \overline{T^*}_\mathrm{bl}, \\
&&&A(\bx) \leq A_\mathrm{bl},  \\
&&&\max_{i,j} \abs{ \theta_{i,j}(\bx) -\tfrac{\pi}{2}}\leq \theta_\mathrm{cr}, \\
&&&\bx \in X,
\end{alignedat}
\end{equation}
where $\eta(\bx)$ is the propulsive efficiency, defined as the ratio of the propulsive power over the aerodynamic power \cite{Stanford2010,Ghommem2012}, $\bx$ is the vector of design parameters describing the perturbations introduced to the locations of the control points,
\[
L^*=\frac{2L}{\rho U_{\infty}^2}, \;T^*=\frac{2T}{\rho U_{\infty}^2}, \;\text{and}\; P^*=\frac{2P}{\rho U_{\infty}^3}
\]
are the normalized lift, thrust, and aerodynamic power, respectively, $A$ is the wing's area, $\rho$ is the fluid density, $U_{\infty}$ is the freestream velocity, the subscript `bl' refers to the baseline case, and the overline denotes a time-averaged quantity over a flapping cycle. The wing geometry is constrained by restricting the feasible perturbations of the locations of the control points to lie in a closed and bounded set $X$. We chose to bound each component in $\bx$ to vary in the interval $[-c/2,c/2]$. To illustrate the wing geometry parametrization based on B-splines, we plot in Figure \ref{Aerogrid} the aerodynamic mesh ($6 \times 10$ lattice) and the location of the control points, represented by red circles, used to define the wing geometry. As pointed out in the previous section, the surface does not interpolate the control points and the aerodynamic mesh does not coincide with the control polygon. Here, the main objective is to maximize the cycle-averaged propulsive efficiency of the wing under lift and thrust constraints. The area of the wing is also restricted to be smaller or equal to a baseline value (i.e., the area of a cambered rectangular wing with an aspect ratio of six). In addition, we require that the angles $\theta_{i,j}$ of each element $i$ where $j=\{1,2,3,4\}$ that correspond to four edges of the element, in the aerodynamic mesh, see Figure \ref{Aerogrid}, do not deviate too much from right angles. Large distortions of the grid could degrade the predictive capability of the aerodynamic model and also may give rise to undesirable unsteady flow effects such as flow separation and substantial wing-wake interactions. In our simulations, we specify the parameter $\theta_\mathrm{cr}$ controlling the level of distortion as $\theta_\mathrm{cr} = 15^\circ$.

The optimization problem is solved with the globally convergent method of moving asymptotes (GCMMA) \cite{Svanberg1987,Svanberg2002}. The algorithm is supplied with numerically-computed gradients based on the first-order backward Euler scheme for both the objective function and the constraints. Several optimization runs are conducted where we vary the number of design variables and the polynomial degree of the basis functions employed in the B-spline representation. These simulations elucidate the effect of the design parameters on the numerical results.

\subsection{Results and discussion}
\subsubsection{Single knot span case}
\begin{figure}[ht]
\begin{center}
\includegraphics[width=0.75\textwidth]{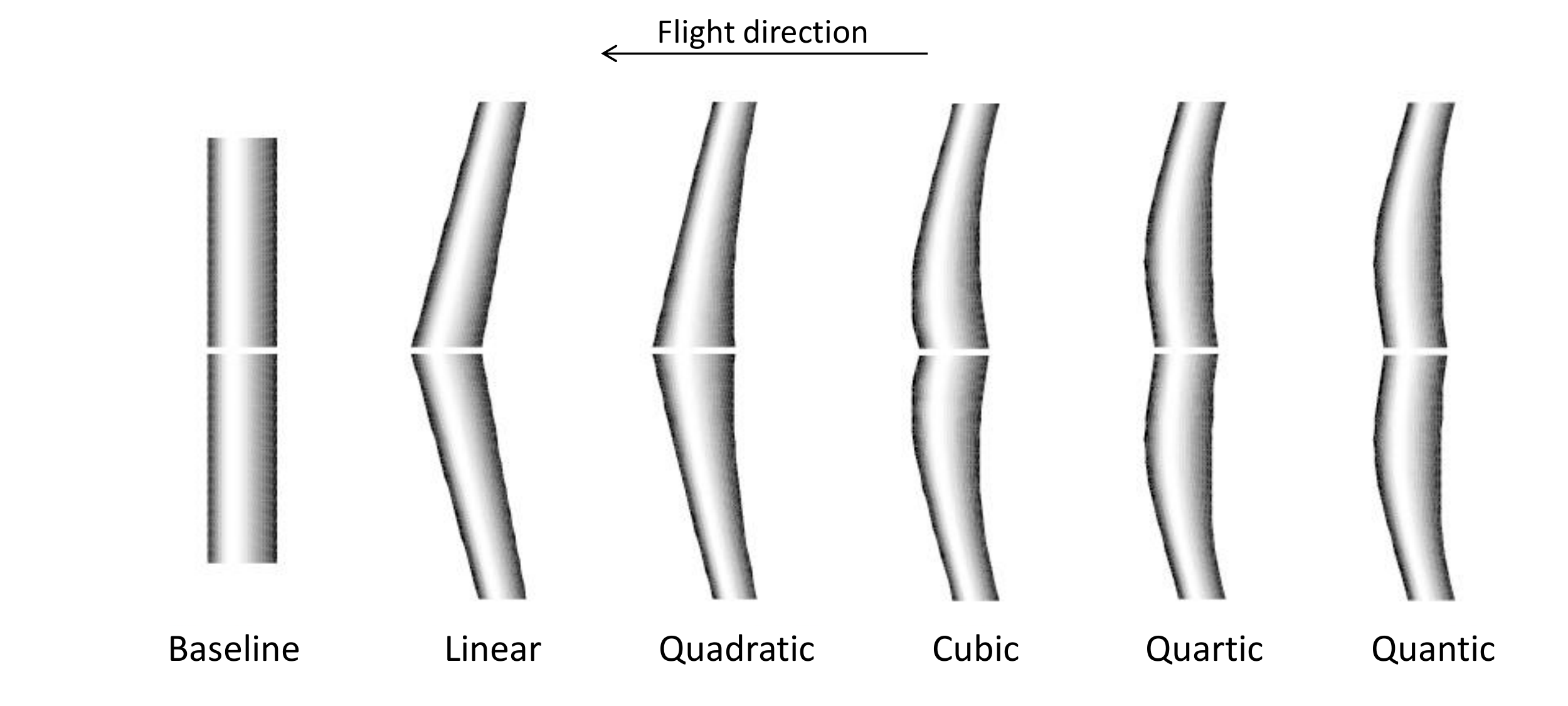}
\end{center}
\caption{Optimal wing shapes (single knot span case). Coordinates of the control points parameterizing the optimal shapes are reported in Tables \ref{Tabs1}--\ref{Tabs4}.} \label{Fig_shape1}
\end{figure}

In the subsequent analysis, we use an aerodynamic mesh of 24 chordwise elements and 20 spanwise elements and the flapping cycle is discretized into 120 time steps. We first consider the case where the B-spline representation is based on a single knot span~\cite{Rogers01,PiegTil97,Farin95,CoRiEl01}. Thus, the number of control points increases with the polynomial degree. The optimal design from one run is used as the initial guess for the next run with a higher polynomial degree. The shapes obtained from the sweep of the optimization studies for different polynomial degrees are shown in Figure~\ref{Fig_shape1}. The control points describing the optimal shapes are listed in Tables \ref{Tabs1}--\ref{Tabs4}. Since the quartic and quintic responses were found similar, the geometric data is not reported for the latter. Optimal results show similar shape trends. For all cases, we observe that less area is distributed to the outer part of the wing. 

\begin{table}
\centering
\caption{Control points for the optimal shape: single knot span case and linear approximation is used in the B-splines representation. Lift, thrust, and area constraints are imposed.}\label{Tabs1}
{\footnotesize
\begin{tabular}{ccccccccccc}
\hline\hline
 & & \multicolumn{3}{c}{Control Points}  & & & & \multicolumn{3}{c}{Control Points}\\
$i$ & $j$ & $x$ & $y$ & $z$ & & $i$ & $j$ & $x$ & $y$ & $z$ \\
\hline\hline
 0 & 0 & -0.015268 & 0.000000 & 0.000000 &  & 1 & 0 & 0.293877 & 3.500000 & 0.000000\\
 0 & 1 & 0.153943 & 0.000000 & 0.089250 &  & 1 & 1 & 0.409020 & 3.500000 & 0.089250\\
 0 & 2 & 0.323154 & 0.000000 & 0.126467 &  & 1 & 2 & 0.524162 & 3.500000 & 0.126467\\
 0 & 3 & 0.492366 & 0.000000 & 0.034380 &  & 1 & 3 & 0.639304 & 3.500000 & 0.034380\\
 0 & 4 & 0.661577 & 0.000000 & 0.091453 &  & 1 & 4 & 0.754447 & 3.500000 & 0.091453\\
 0 & 5 & 0.830789 & 0.000000 & 0.031200 &  & 1 & 5 & 0.869589 & 3.500000 & 0.031200\\
 0 & 6 & 1.000000 & 0.000000 & 0.000000 &  & 1 & 6 & 0.984732 & 3.500000 & 0.000000\\
\hline\hline
\end{tabular}}
\end{table}
\begin{table}
\centering
\caption{Control points for the optimal shape: single knot span case and quadratic approximation is used in the B-splines representation. Lift, thrust, and area constraints are imposed.}\label{Tabs2}
{\footnotesize
\begin{tabular}{ccccccccccc}
\hline\hline
 & & \multicolumn{3}{c}{Control Points}  & & & & \multicolumn{3}{c}{Control Points}\\
$i$ & $j$ & $x$ & $y$ & $z$ & & $i$ & $j$ & $x$ & $y$ & $z$ \\
\hline\hline
 0 & 0 & -0.238849 & 0.000000 & 0.000000 &  & 2 & 0 & 0.740752 & 3.500000 & 0.000000\\
 0 & 1 & -0.032374 & 0.000000 & 0.089250 &  & 2 & 1 & 0.834122 & 3.500000 & 0.089250\\
 0 & 2 & 0.174101 & 0.000000 & 0.126467 &  & 2 & 2 & 0.927493 & 3.500000 & 0.126467\\
 0 & 3 & 0.380576 & 0.000000 & 0.034380 &  & 2 & 3 & 1.020860 & 3.500000 & 0.034380\\
 0 & 4 & 0.587050 & 0.000000 & 0.091453 &  & 2 & 4 & 1.114230 & 3.500000 & 0.091453\\
 0 & 5 & 0.793525 & 0.000000 & 0.031200 &  & 2 & 5 & 1.207600 & 3.500000 & 0.031200\\
 0 & 6 & 1.000000 & 0.000000 & 0.000000 &  & 2 & 6 & 1.300970 & 3.500000 & 0.000000\\
 1 & 0 & 0.176031 & 1.500000 & 0.000000 &  & . & . & . & . & . \\
 1 & 1 & 0.301366 & 1.500000 & 0.089250 &  & . & . & . & . & . \\
 1 & 2 & 0.426702 & 1.500000 & 0.126467 &  & . & . & . & . & . \\
 1 & 3 & 0.552037 & 1.500000 & 0.034380 &  & . & . & . & . & . \\
 1 & 4 & 0.677372 & 1.500000 & 0.091453 &  & . & . & . & . & . \\
 1 & 5 & 0.802708 & 1.500000 & 0.031200 &  & . & . & . & . & . \\
 1 & 6 & 0.928043 & 1.500000 & 0.000000 &  & . & . & . & . & . \\
\hline\hline
\end{tabular}}
\end{table}
\begin{table}
\centering
\caption{Control points for the optimal shape: single knot span case and cubic approximation is used in the B-splines representation. Lift, thrust, and area constraints are imposed.}\label{Tabs3}
{\footnotesize
\begin{tabular}{ccccccccccc}
\hline\hline
 & & \multicolumn{3}{c}{Control Points}  & & & & \multicolumn{3}{c}{Control Points}\\
$i$ & $j$ & $x$ & $y$ & $z$ & & $i$ & $j$ & $x$ & $y$ & $z$ \\
\hline\hline
 0 & 0 & 0.004282 & 0.000000 & 0.000000 &  & 2 & 0 & 0.112514 & 2.000000 & 0.000000\\
 0 & 1 & 0.170235 & 0.000000 & 0.089250 &  & 2 & 1 & 0.219737 & 2.000000 & 0.089250\\
 0 & 2 & 0.336188 & 0.000000 & 0.126467 &  & 2 & 2 & 0.326960 & 2.000000 & 0.126467\\
 0 & 3 & 0.502141 & 0.000000 & 0.034380 &  & 2 & 3 & 0.434183 & 2.000000 & 0.034380\\
 0 & 4 & 0.668094 & 0.000000 & 0.091453 &  & 2 & 4 & 0.541405 & 2.000000 & 0.091453\\
 0 & 5 & 0.834047 & 0.000000 & 0.031200 &  & 2 & 5 & 0.648628 & 2.000000 & 0.031200\\
 0 & 6 & 1.000000 & 0.000000 & 0.000000 &  & 2 & 6 & 0.755851 & 2.000000 & 0.000000\\
 1 & 0 & -0.315068 & 1.000000 & 0.000000 &  & 3 & 0 & 0.491455 & 3.500000 & 0.000000\\
 1 & 1 & -0.119715 & 1.000000 & 0.089250 &  & 3 & 1 & 0.600877 & 3.500000 & 0.089250\\
 1 & 2 & 0.075637 & 1.000000 & 0.126467 &  & 3 & 2 & 0.710300 & 3.500000 & 0.126467\\
 1 & 3 & 0.270989 & 1.000000 & 0.034380 &  & 3 & 3 & 0.819722 & 3.500000 & 0.034380\\
 1 & 4 & 0.466341 & 1.000000 & 0.091453 &  & 3 & 4 & 0.929145 & 3.500000 & 0.091453\\
 1 & 5 & 0.661693 & 1.000000 & 0.031200 &  & 3 & 5 & 1.038570 & 3.500000 & 0.031200\\
 1 & 6 & 0.857045 & 1.000000 & 0.000000 &  & 3 & 6 & 1.147990 & 3.500000 & 0.000000\\
\hline\hline
\end{tabular}}
\end{table}
\begin{table}
\centering
\caption{Control points for the optimal shape: single knot span case and quartic approximation is used in the B-splines representation. Lift, thrust, and area constraints are imposed.}\label{Tabs4}
{\footnotesize
\begin{tabular}{ccccccccccc}
\hline\hline
 & & \multicolumn{3}{c}{Control Points}  & & & & \multicolumn{3}{c}{Control Points}\\
$i$ & $j$ & $x$ & $y$ & $z$ & & $i$ & $j$ & $x$ & $y$ & $z$ \\
\hline\hline
 0 & 0 & 0.089656 & 0.000000 & 0.000000 &  & 3 & 0 & 0.149637 & 2.250000 & 0.000000\\
 0 & 1 & 0.241380 & 0.000000 & 0.089250 &  & 3 & 1 & 0.249139 & 2.250000 & 0.089250\\
 0 & 2 & 0.393104 & 0.000000 & 0.126467 &  & 3 & 2 & 0.348641 & 2.250000 & 0.126467\\
 0 & 3 & 0.544828 & 0.000000 & 0.034380 &  & 3 & 3 & 0.448142 & 2.250000 & 0.034380\\
 0 & 4 & 0.696552 & 0.000000 & 0.091453 &  & 3 & 4 & 0.547644 & 2.250000 & 0.091453\\
 0 & 5 & 0.848276 & 0.000000 & 0.031200 &  & 3 & 5 & 0.647145 & 2.250000 & 0.031200\\
 0 & 6 & 1.000000 & 0.000000 & 0.000000 &  & 3 & 6 & 0.746647 & 2.250000 & 0.000000\\
 1 & 0 & 0.026628 & 0.750000 & 0.000000 &  & 4 & 0 & 0.444682 & 3.500000 & 0.000000\\
 1 & 1 & 0.165262 & 0.750000 & 0.089250 &  & 4 & 1 & 0.555822 & 3.500000 & 0.089250\\
 1 & 2 & 0.303896 & 0.750000 & 0.126467 &  & 4 & 2 & 0.666962 & 3.500000 & 0.126467\\
 1 & 3 & 0.442530 & 0.750000 & 0.034380 &  & 4 & 3 & 0.778102 & 3.500000 & 0.034380\\
 1 & 4 & 0.581163 & 0.750000 & 0.091453 &  & 4 & 4 & 0.889242 & 3.500000 & 0.091453\\
 1 & 5 & 0.719797 & 0.750000 & 0.031200 &  & 4 & 5 & 1.000380 & 3.500000 & 0.031200\\
 1 & 6 & 0.858431 & 0.750000 & 0.000000 &  & 4 & 6 & 1.111520 & 3.500000 & 0.000000\\
 2 & 0 & -0.292502 & 1.500000 & 0.000000 &  & . & . & . & . & . \\
 2 & 1 & -0.071887 & 1.500000 & 0.089250 &  & . & . & . & . & . \\
 2 & 2 & 0.148727 & 1.500000 & 0.126467 &  & . & . & . & . & . \\
 2 & 3 & 0.369341 & 1.500000 & 0.034380 &  & . & . & . & . & . \\
 2 & 4 & 0.589956 & 1.500000 & 0.091453 &  & . & . & . & . & . \\
 2 & 5 & 0.810570 & 1.500000 & 0.031200 &  & . & . & . & . & . \\
 2 & 6 & 1.031180 & 1.500000 & 0.000000 &  & . & . & . & . & . \\
\hline\hline
\end{tabular}}
\end{table}

Table \ref{OPTk01SE} provides a summary of the efficiency $\eta$, lift $\overline{L^*}$, thrust $\overline{T^*}$, aerodynamic power $\overline{P^*}$, and area ratio $A/A_\mathrm{bl}$ obtained for the optimal configurations for different polynomial degrees. The number of design variables is denoted by $N_\mathrm{DV}$. The optimal results show that changing the wing shape allows to achieve higher lift and propulsive efficiency but also more power is needed to be introduced to the flying system. As expected, rising the polynomial degree (i.e., increasing the number of degrees of freedom of the parametrization $N_\mathrm{DV}$) yields optimal shapes that enable flapping flights with higher efficiencies. Introducing swept and taper to flapping wings, which is obtained by using of linear polynomials in the B-spline representation, ameliorates significantly the flight performance. The use of higher polynomial degrees enables further improvements in the propulsive efficiency which stabilizes at fourth degree. Furthermore, we observe that the area constraint is active for all cases. Thus, there is a natural trade-off between the wing area and efficiency, but not so for thrust and efficiency. This indicates that, in the optimized configuration obtained at the low reduced frequency ($\kappa = 0.1$), changing the wing shape enables to improve thrust generation without affecting the wing area.
\begin{table}[ht!]
\caption{Baseline vs. optimal results obtained for the single knot span case ($\kappa = 0.1$). Lift, thrust, and area constraints are imposed.}\label{OPTk01SE}
\begin{center}
\begin{tabular}{l l l l l l l l}
  \hline
  \hline \\[-13pt]
  Wing shape & & $N_\mathrm{DV}$ & $\eta$ & $\overline{L^*}$ & $\overline{T^*}$ & $\overline{P^*}$ & $A/A_\mathrm{bl}$ \\
  \hline
  \hline
  Baseline shape & & 0 &0.219 & 4.484 & 0.185 & 0.844 & 1\\
  \hline
  \hline
  Optimal & Linear & 4 & 0.426 & 5.421 & 0.4451 & 1.0435 & 1 \\
   & Quadratic & 6 & 0.433 & 5.448 & 0.415 & 0.9566 & 1 \\
   & Cubic & 8 & 0.439 & 5.489 & 0.443 & 1.009 & 1 \\
   \rowcolor[gray]{.9} & Quartic & 10 & 0.440 & 5.470 & 0.459 & 1.040 & 1 \\
   \rowcolor[gray]{.9} & Quintic & 12 & 0.440 & 5.470 & 0.459 & 1.040 & 1 \\
  \hline
  \hline
\end{tabular}
\end{center}
\end{table}

\subsubsection{Fixed number of degrees of freedom case}\label{FDOF}
\begin{figure}[ht]
\begin{center}
\includegraphics[width=0.75\textwidth]{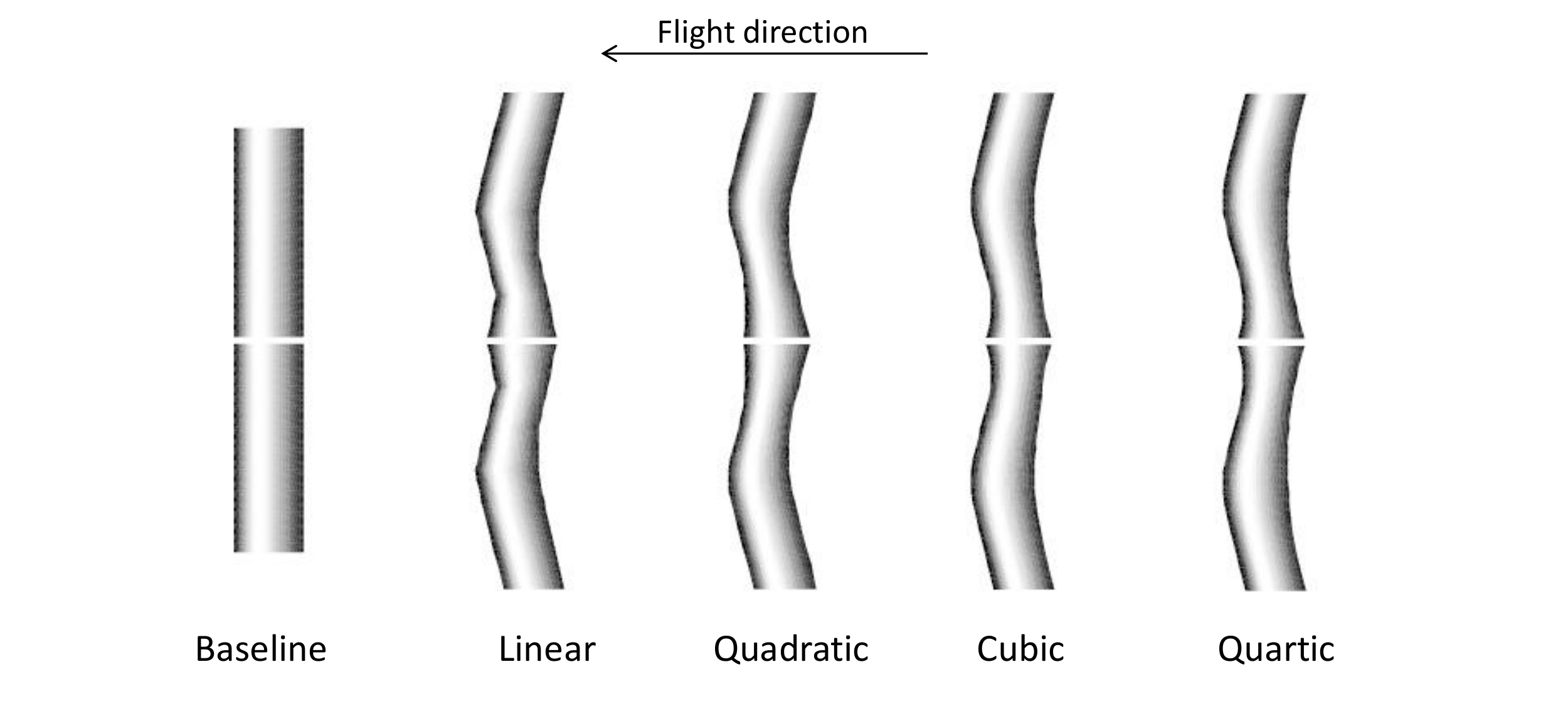}
\end{center}
\caption{Optimal wing shapes (low reduced frequency, $\kappa=0.1$, and fixed number of degrees of freedom case, $N_\mathrm{DV}=12$ ). Coordinates of the control points parameterizing the optimal shapes are reported in Tables \ref{Tabs5}--\ref{Tabs7}.} \label{Fig_shape2}
\end{figure}

In this case, we keep the number of design variables (degrees of freedom) fixed and equal to $N_\mathrm{DV}=12$ and run the optimizer while varying the polynomial degree using in all instances the baseline configuration as the optimization starting point. The choice of $N_\mathrm{DV}$ is made so that it coincides with the number of design variables required for the single knot span case when considering a fifth-degree polynomial. The corresponding optimal shapes are shown in Figure~\ref{Fig_shape2}. The control points describing the optimal shapes are listed in Tables \ref{Tabs5}--\ref{Tabs7}. Again, since the cubic and quartic responses were found similar, the geometric data is not reported for the latter. Figure~\ref{Fig_shape2} shows that all optimal shapes are qualitatively similar. In particular, the ones obtained using higher-degree polynomials look smoother. This smoothness in the shape affects the performance of the flapping wing. Although the single knot span case using fifth-degree polynomial and the fixed number of degrees of freedom cases have the same number of design variables, different shapes have been obtained. This difference can be associated to the choice of the starting shape when running the optimizer. Furthermore, we observe that the curvatures of the leading and trailing edges are both tilted up near the wing tip. Several studies have pointed out the importance of the curvature of the leading edge as being the location of the first contact of the fluid with the body and the camber line~\cite{Leung2012,Lian2004}.

\begin{table}
\centering
\caption{Control points for the optimal shape: fixed number of degrees of freedom case and linear approximation is used in the B-splines representation. Lift, thrust, and area constraints are imposed.}\label{Tabs5}
{\footnotesize
\begin{tabular}{ccccccccccc}
\hline\hline
 & & \multicolumn{3}{c}{Control Points}  & & & & \multicolumn{3}{c}{Control Points}\\
$i$ & $j$ & $x$ & $y$ & $z$ & & $i$ & $j$ & $x$ & $y$ & $z$ \\
\hline\hline
 0 & 0 & 0.019323 & 0.000000 & 0.000000 &  & 3 & 0 & -0.160116 & 1.800000 & 0.000000\\
 0 & 1 & 0.182769 & 0.000000 & 0.089250 &  & 3 & 1 & -0.009644 & 1.800000 & 0.089250\\
 0 & 2 & 0.346216 & 0.000000 & 0.126467 &  & 3 & 2 & 0.140827 & 1.800000 & 0.126467\\
 0 & 3 & 0.509662 & 0.000000 & 0.034380 &  & 3 & 3 & 0.291299 & 1.800000 & 0.034380\\
 0 & 4 & 0.673108 & 0.000000 & 0.091453 &  & 3 & 4 & 0.441771 & 1.800000 & 0.091453\\
 0 & 5 & 0.836554 & 0.000000 & 0.031200 &  & 3 & 5 & 0.592243 & 1.800000 & 0.031200\\
 0 & 6 & 1.000000 & 0.000000 & 0.000000 &  & 3 & 6 & 0.742714 & 1.800000 & 0.000000\\
 1 & 0 & 0.136362 & 0.600000 & 0.000000 &  & 4 & 0 & -0.040048 & 2.400000 & 0.000000\\
 1 & 1 & 0.260379 & 0.600000 & 0.089250 &  & 4 & 1 & 0.108170 & 2.400000 & 0.089250\\
 1 & 2 & 0.384396 & 0.600000 & 0.126467 &  & 4 & 2 & 0.256389 & 2.400000 & 0.126467\\
 1 & 3 & 0.508413 & 0.600000 & 0.034380 &  & 4 & 3 & 0.404607 & 2.400000 & 0.034380\\
 1 & 4 & 0.632430 & 0.600000 & 0.091453 &  & 4 & 4 & 0.552825 & 2.400000 & 0.091453\\
 1 & 5 & 0.756447 & 0.600000 & 0.031200 &  & 4 & 5 & 0.701043 & 2.400000 & 0.031200\\
 1 & 6 & 0.880463 & 0.600000 & 0.000000 &  & 4 & 6 & 0.849262 & 2.400000 & 0.000000\\
 2 & 0 & -0.015657 & 1.200000 & 0.000000 &  & 5 & 0 & 0.244208 & 3.500000 & 0.000000\\
 2 & 1 & 0.112085 & 1.200000 & 0.089250 &  & 5 & 1 & 0.389343 & 3.500000 & 0.089250\\
 2 & 2 & 0.239827 & 1.200000 & 0.126467 &  & 5 & 2 & 0.534479 & 3.500000 & 0.126467\\
 2 & 3 & 0.367569 & 1.200000 & 0.034380 &  & 5 & 3 & 0.679614 & 3.500000 & 0.034380\\
 2 & 4 & 0.495311 & 1.200000 & 0.091453 &  & 5 & 4 & 0.824750 & 3.500000 & 0.091453\\
 2 & 5 & 0.623053 & 1.200000 & 0.031200 &  & 5 & 5 & 0.969885 & 3.500000 & 0.031200\\
 2 & 6 & 0.750795 & 1.200000 & 0.000000 &  & 5 & 6 & 1.115020 & 3.500000 & 0.000000\\
\hline\hline
\end{tabular}}
\end{table}
\begin{table}
\centering
\caption{Control points for the optimal shape: fixed number of degrees of freedom case and quadratic approximation is used in the B-splines representation. Lift, thrust, and area constraints are imposed.}\label{Tabs6}
{\footnotesize
\begin{tabular}{ccccccccccc}
\hline\hline
 & & \multicolumn{3}{c}{Control Points}  & & & & \multicolumn{3}{c}{Control Points}\\
$i$ & $j$ & $x$ & $y$ & $z$ & & $i$ & $j$ & $x$ & $y$ & $z$ \\
\hline\hline
 0 & 0 & 0.060584 & 0.000000 & 0.000000 &  & 3 & 0 & -0.223082 & 1.875000 & 0.000000\\
 0 & 1 & 0.217153 & 0.000000 & 0.089250 &  & 3 & 1 & -0.071007 & 1.875000 & 0.089250\\
 0 & 2 & 0.373723 & 0.000000 & 0.126467 &  & 3 & 2 & 0.081068 & 1.875000 & 0.126467\\
 0 & 3 & 0.530292 & 0.000000 & 0.034380 &  & 3 & 3 & 0.233144 & 1.875000 & 0.034380\\
 0 & 4 & 0.686861 & 0.000000 & 0.091453 &  & 3 & 4 & 0.385219 & 1.875000 & 0.091453\\
 0 & 5 & 0.843431 & 0.000000 & 0.031200 &  & 3 & 5 & 0.537294 & 1.875000 & 0.031200\\
 0 & 6 & 1.000000 & 0.000000 & 0.000000 &  & 3 & 6 & 0.689369 & 1.875000 & 0.000000\\
 1 & 0 & 0.103366 & 0.375000 & 0.000000 &  & 4 & 0 & -0.034651 & 2.625000 & 0.000000\\
 1 & 1 & 0.236516 & 0.375000 & 0.089250 &  & 4 & 1 & 0.119043 & 2.625000 & 0.089250\\
 1 & 2 & 0.369666 & 0.375000 & 0.126467 &  & 4 & 2 & 0.272736 & 2.625000 & 0.126467\\
 1 & 3 & 0.502816 & 0.375000 & 0.034380 &  & 4 & 3 & 0.426430 & 2.625000 & 0.034380\\
 1 & 4 & 0.635966 & 0.375000 & 0.091453 &  & 4 & 4 & 0.580123 & 2.625000 & 0.091453\\
 1 & 5 & 0.769116 & 0.375000 & 0.031200 &  & 4 & 5 & 0.733817 & 2.625000 & 0.031200\\
 1 & 6 & 0.902266 & 0.375000 & 0.000000 &  & 4 & 6 & 0.887511 & 2.625000 & 0.000000\\
 2 & 0 & 0.031374 & 1.125000 & 0.000000 &  & 5 & 0 & 0.191797 & 3.500000 & 0.000000\\
 2 & 1 & 0.145334 & 1.125000 & 0.089250 &  & 5 & 1 & 0.344134 & 3.500000 & 0.089250\\
 2 & 2 & 0.259295 & 1.125000 & 0.126467 &  & 5 & 2 & 0.496471 & 3.500000 & 0.126467\\
 2 & 3 & 0.373256 & 1.125000 & 0.034380 &  & 5 & 3 & 0.648809 & 3.500000 & 0.034380\\
 2 & 4 & 0.487216 & 1.125000 & 0.091453 &  & 5 & 4 & 0.801146 & 3.500000 & 0.091453\\
 2 & 5 & 0.601177 & 1.125000 & 0.031200 &  & 5 & 5 & 0.953483 & 3.500000 & 0.031200\\
 2 & 6 & 0.715137 & 1.125000 & 0.000000 &  & 5 & 6 & 1.105820 & 3.500000 & 0.000000\\
\hline\hline
\end{tabular}}
\end{table}
\begin{table}
\centering
\caption{Control points for the optimal shape: fixed number of degrees of freedom case and cubic approximation is used in the B-splines representation. Lift, thrust, and area constraints are imposed.}\label{Tabs7}
{\footnotesize
\begin{tabular}{ccccccccccc}
\hline\hline
 & & \multicolumn{3}{c}{Control Points}  & & & & \multicolumn{3}{c}{Control Points}\\
$i$ & $j$ & $x$ & $y$ & $z$ & & $i$ & $j$ & $x$ & $y$ & $z$ \\
\hline\hline
 0 & 0 & 0.181755 & 0.000000 & 0.000000 &  & 3 & 0 & -0.261467 & 2.000000 & 0.000000\\
 0 & 1 & 0.318130 & 0.000000 & 0.089250 &  & 3 & 1 & -0.099029 & 2.000000 & 0.089250\\
 0 & 2 & 0.454504 & 0.000000 & 0.126467 &  & 3 & 2 & 0.063409 & 2.000000 & 0.126467\\
 0 & 3 & 0.590878 & 0.000000 & 0.034380 &  & 3 & 3 & 0.225847 & 2.000000 & 0.034380\\
 0 & 4 & 0.727252 & 0.000000 & 0.091453 &  & 3 & 4 & 0.388285 & 2.000000 & 0.091453\\
 0 & 5 & 0.863626 & 0.000000 & 0.031200 &  & 3 & 5 & 0.550723 & 2.000000 & 0.031200\\
 0 & 6 & 1.000000 & 0.000000 & 0.000000 &  & 3 & 6 & 0.713161 & 2.000000 & 0.000000\\
 1 & 0 & 0.240014 & 0.333333 & 0.000000 &  & 4 & 0 & -0.009370 & 2.666670 & 0.000000\\
 1 & 1 & 0.359515 & 0.333333 & 0.089250 &  & 4 & 1 & 0.148156 & 2.666670 & 0.089250\\
 1 & 2 & 0.479017 & 0.333333 & 0.126467 &  & 4 & 2 & 0.305682 & 2.666670 & 0.126467\\
 1 & 3 & 0.598518 & 0.333333 & 0.034380 &  & 4 & 3 & 0.463208 & 2.666670 & 0.034380\\
 1 & 4 & 0.718020 & 0.333333 & 0.091453 &  & 4 & 4 & 0.620734 & 2.666670 & 0.091453\\
 1 & 5 & 0.837521 & 0.333333 & 0.031200 &  & 4 & 5 & 0.778260 & 2.666670 & 0.031200\\
 1 & 6 & 0.957023 & 0.333333 & 0.000000 &  & 4 & 6 & 0.935786 & 2.666670 & 0.000000\\
 2 & 0 & 0.138733 & 1.000000 & 0.000000 &  & 5 & 0 & 0.198005 & 3.500000 & 0.000000\\
 2 & 1 & 0.250931 & 1.000000 & 0.089250 &  & 5 & 1 & 0.353562 & 3.500000 & 0.089250\\
 2 & 2 & 0.363129 & 1.000000 & 0.126467 &  & 5 & 2 & 0.509119 & 3.500000 & 0.126467\\
 2 & 3 & 0.475326 & 1.000000 & 0.034380 &  & 5 & 3 & 0.664675 & 3.500000 & 0.034380\\
 2 & 4 & 0.587524 & 1.000000 & 0.091453 &  & 5 & 4 & 0.820232 & 3.500000 & 0.091453\\
 2 & 5 & 0.699721 & 1.000000 & 0.031200 &  & 5 & 5 & 0.975788 & 3.500000 & 0.031200\\
 2 & 6 & 0.811919 & 1.000000 & 0.000000 &  & 5 & 6 & 1.131340 & 3.500000 & 0.000000\\
\hline\hline
\end{tabular}}
\end{table}

 Table \ref{OPTk01} provides a summary of the efficiency $\eta$, lift $\overline{L^*}$, thrust $\overline{T^*}$, aerodynamic power $\overline{P^*}$, and area ratio $A/A_\mathrm{bl}$ obtained for the optimal configurations for different polynomials. Clearly, linear polynomials do a good job in improving the performance of flapping wings. The optimal results show mild changes when considering higher-degree polynomials and converge at the third degree. The optimal shape yields an increase in the lift while the overall wing area was kept the same as the baseline case. The required input power increases as well however. This indicates that not only changing the wing shape allows to achieve an efficiency of $\eta=0.441$ ($\sim 100 \%$ increase) but also more power needs to be introduced to the flying system. The set of optimal shapes obtained for the present case enables generation of higher thrust and lower lift but requires more aerodynamic power while maintaining almost the same level of efficiency in comparison with the single knot span case. This observation can be related to the wing area distribution. In fact, the proportion of area in the outer part of the wing (near the tip) which undergoes higher acceleration is thicker and then it needs more power to move. Besides, we remark that the optimal wings are longer (i.e., higher aspect ratio). The optimal wing tip, which is subjected to higher angular acceleration, is pushed further away. This explains the significant increase in the aerodynamic power with respect the baseline case. These observations provide insights on how to design efficient wing planforms based on the mission requirements. As such, shortening the wing tip helps in increasing the lift force and decreasing the required aerodynamic power. This feature can be useful for take-off flights. Furthermore, the differences observed in the shapes obtained from the different optimization cases explains the diversity of wing morphologies observed in nature. 

\begin{table}[ht!]
\caption{Baseline vs. optimal results obtained for the fixed number of degrees of freedom case ($\kappa=0.1$). Lift, thrust, and area constraints are imposed.}\label{OPTk01}
\begin{center}
\begin{tabular}{l l l l l l l}
  \hline
  \hline \\[-13pt]
  Wing shape & & $\eta$ & $\overline{L^*}$ & $\overline{T^*}$ & $\overline{P^*}$ & $A/A_\mathrm{bl}$ \\
  \hline
  Baseline shape & -- &0.219 & 4.484 & 0.185 & 0.844 & 1 \\
  \hline
  Optimal shapes & Linear & 0.439 & 5.390 & 0.498 & 1.134 & 1 \\
   & Quadratic & 0.440 & 5.374 & 0.506 & 1.148 & 1 \\
   \rowcolor[gray]{.9} & Cubic & 0.441 & 5.398 & 0.506 & 1.146 & 1 \\
   \rowcolor[gray]{.9} & Quartic & 0.441 & 5.398 & 0.506 & 1.146 & 1 \\
  \hline
  \hline
\end{tabular}
\end{center}
\end{table}

\begin{figure}
  \begin{center}
      \subfigure[Lift]{\includegraphics[width=0.45\textwidth]{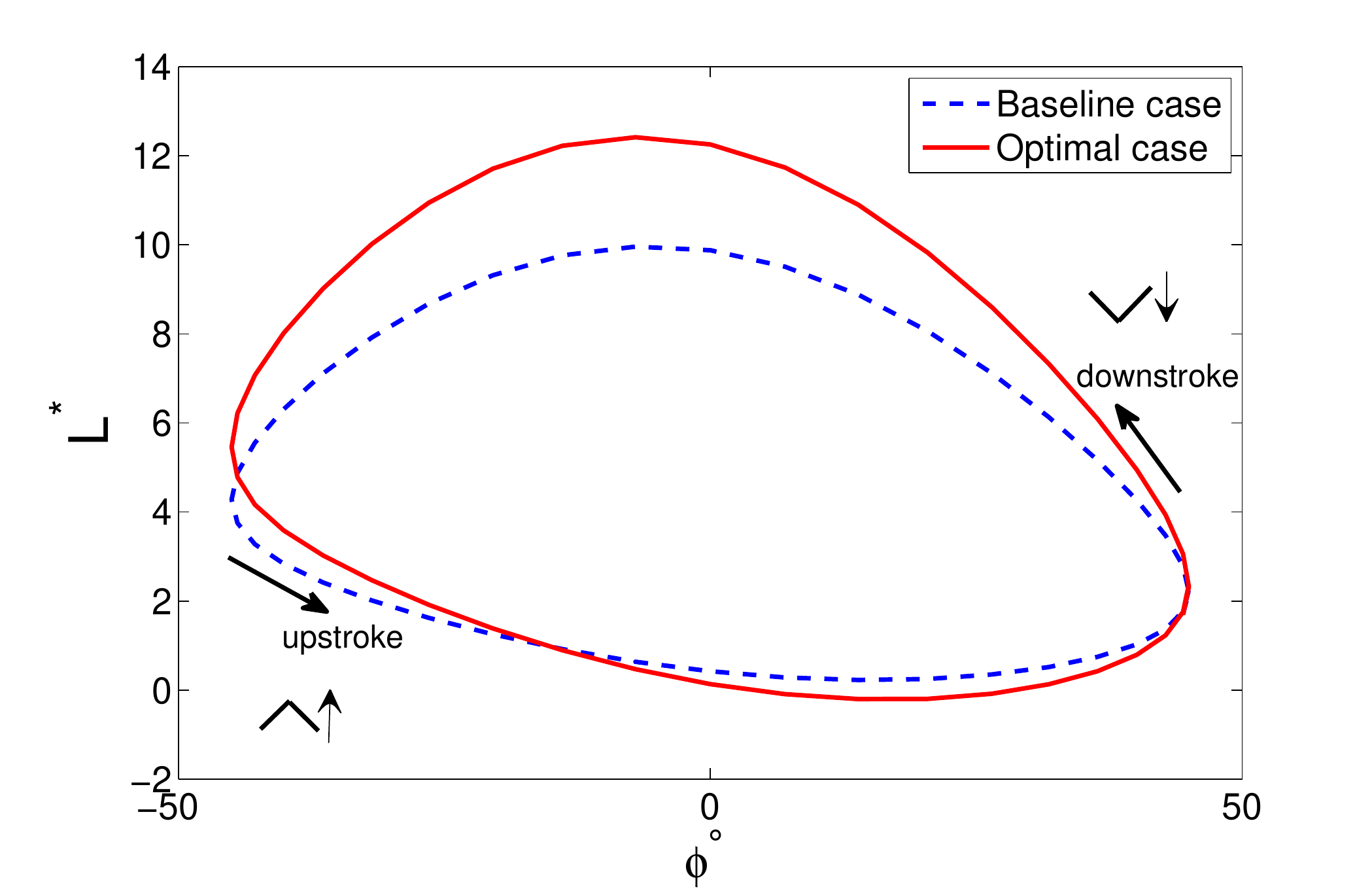}\label{Fig4a}}
      \subfigure[Thrust]{\includegraphics[width=0.45\textwidth]{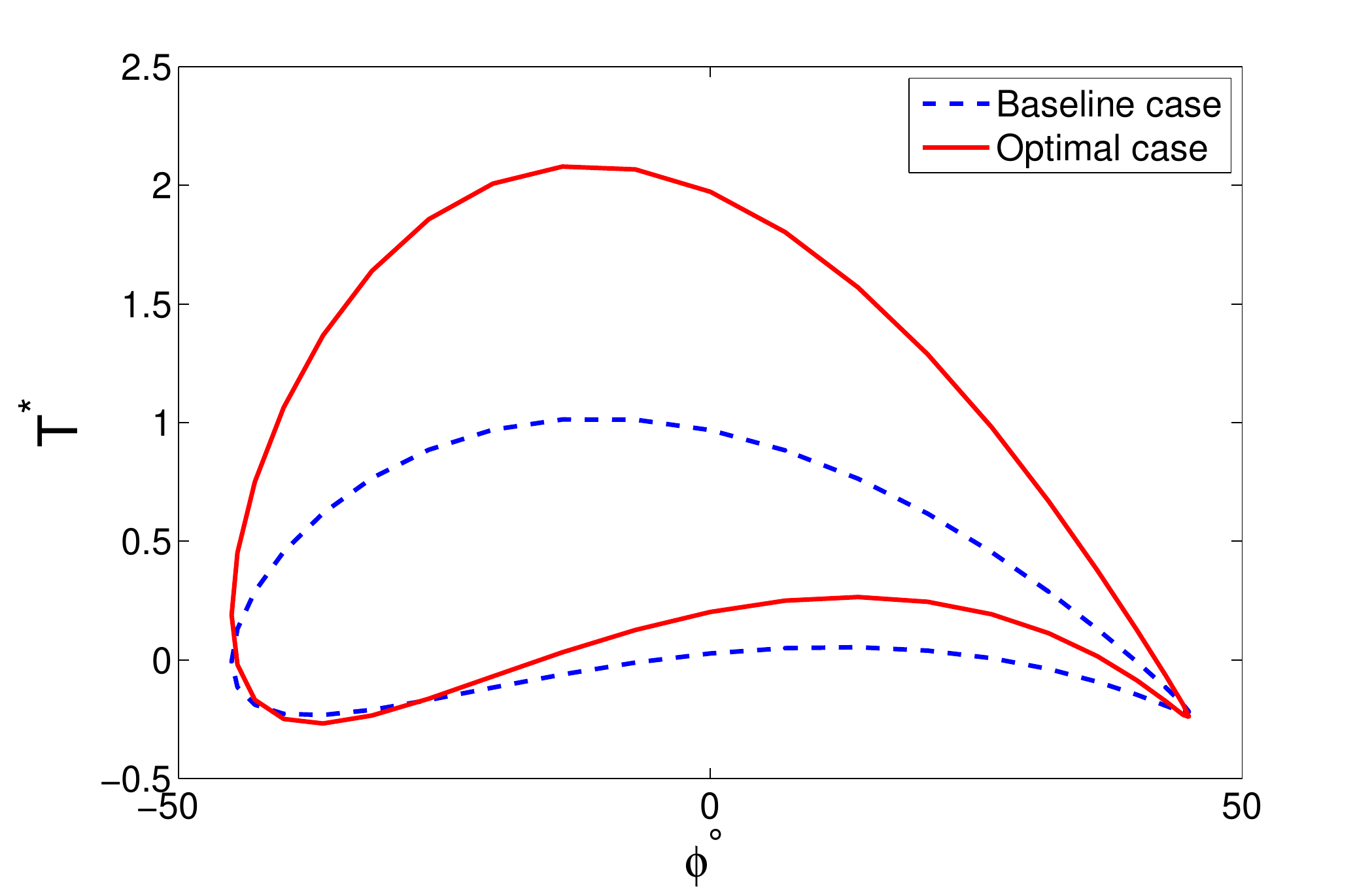}\label{Fig4b}}
      \subfigure[Power]{\includegraphics[width=0.45\textwidth]{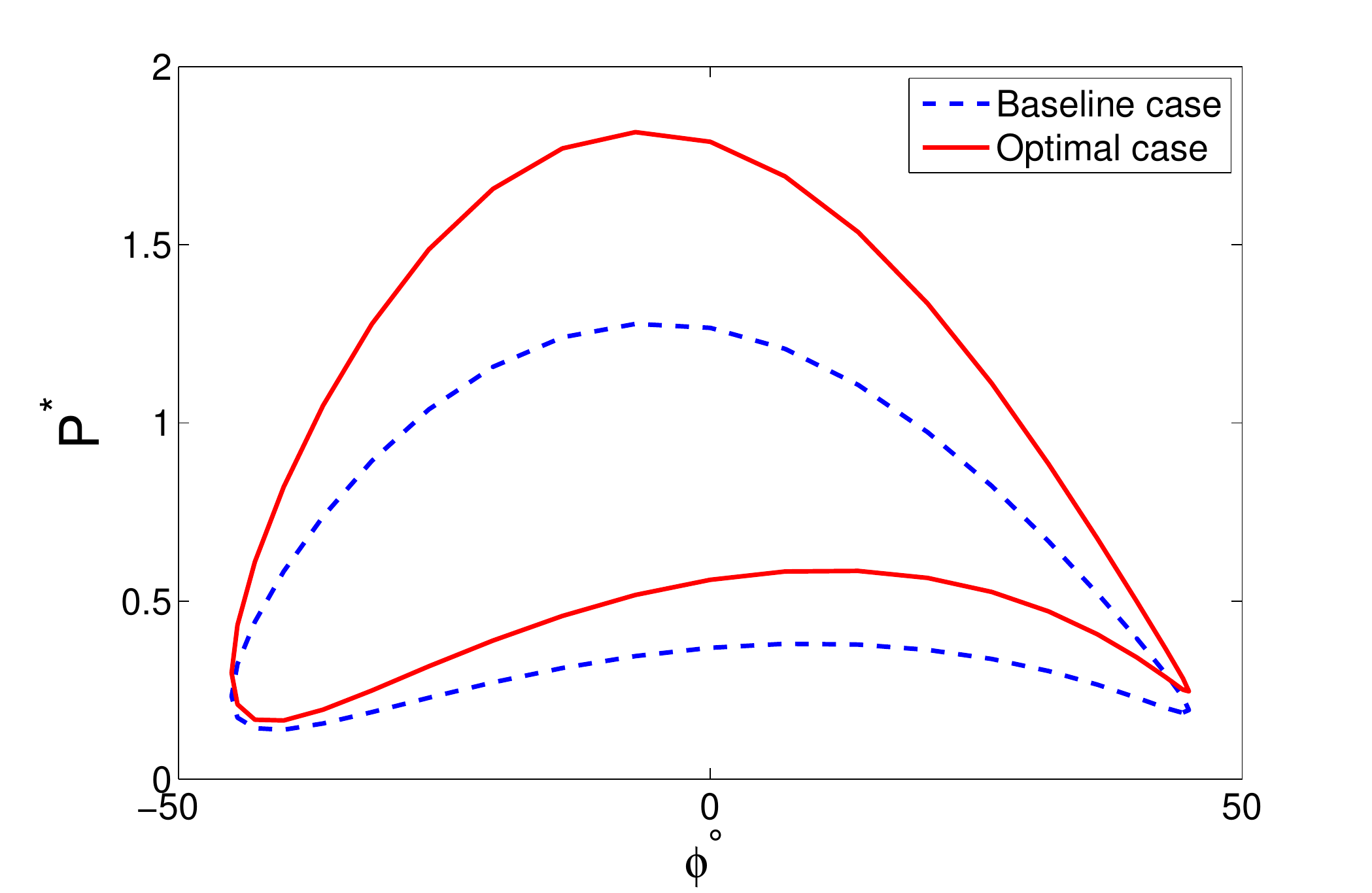}\label{Fig4c}}
  \end{center}
  \caption{Lift, thrust, and power plotted versus the flapping angle $\phi$ for the baseline (rectangular) and optimal shapes ($\kappa=0.1$).}
  \label{Fig4}
\end{figure}

Figure~\ref{Fig4} shows the lift, thrust, and aerodynamic power that develop over the flapping wing at $\kappa=0.1$ for both the baseline and optimal cases, plotted as function of the flapping angle $\phi$. As expected, the bulk of the useful aerodynamic forces (positive lift and thrust) are generated during the downstroke. In particular, thrust and lift peaks are reached near the middle of the downstroke phase. Positive lift is produced during both strokes, as the angle of attack induced by the flapping motion is smaller than the fixed pitch angle (5$^{\circ}$) at the wing root, and thus the lift remains positive through almost the entire cycle. The optimal shape does not alter the phases of the aerodynamic quantities, since the effective angle of attack does not vary significantly over the flapping cycle. Nevertheless, it is able to increase the time-averaged lift, thrust, and power (as seen in Table \ref{OPTk01}), and also their peaks as shown in Figure~\ref{Fig4}.

\begin{figure}[ht]
  \begin{center}
      \subfigure[Baseline case]{\includegraphics[width=0.48\textwidth]{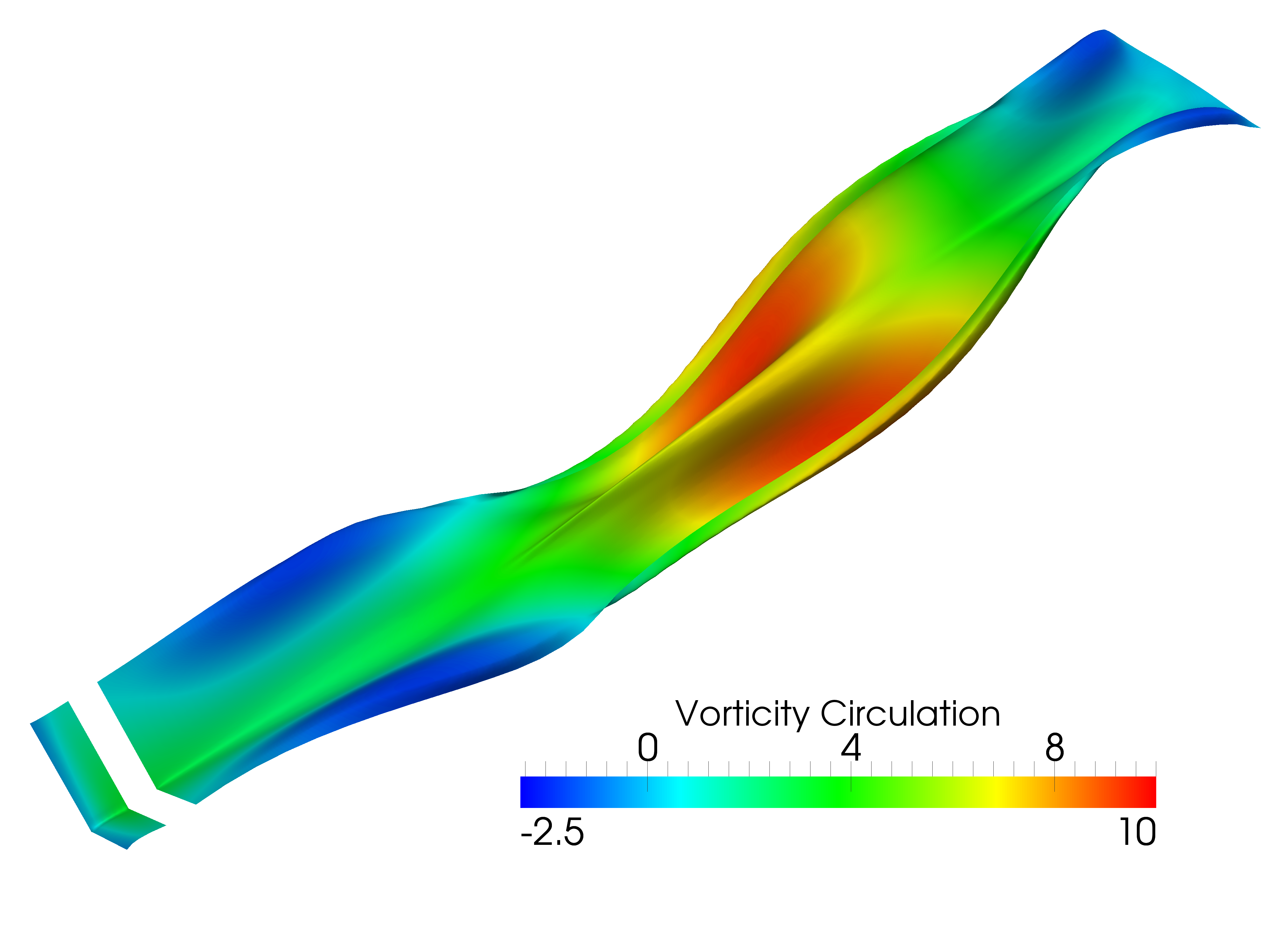}}
      \subfigure[Optimal case]{\includegraphics[width=0.48\textwidth]{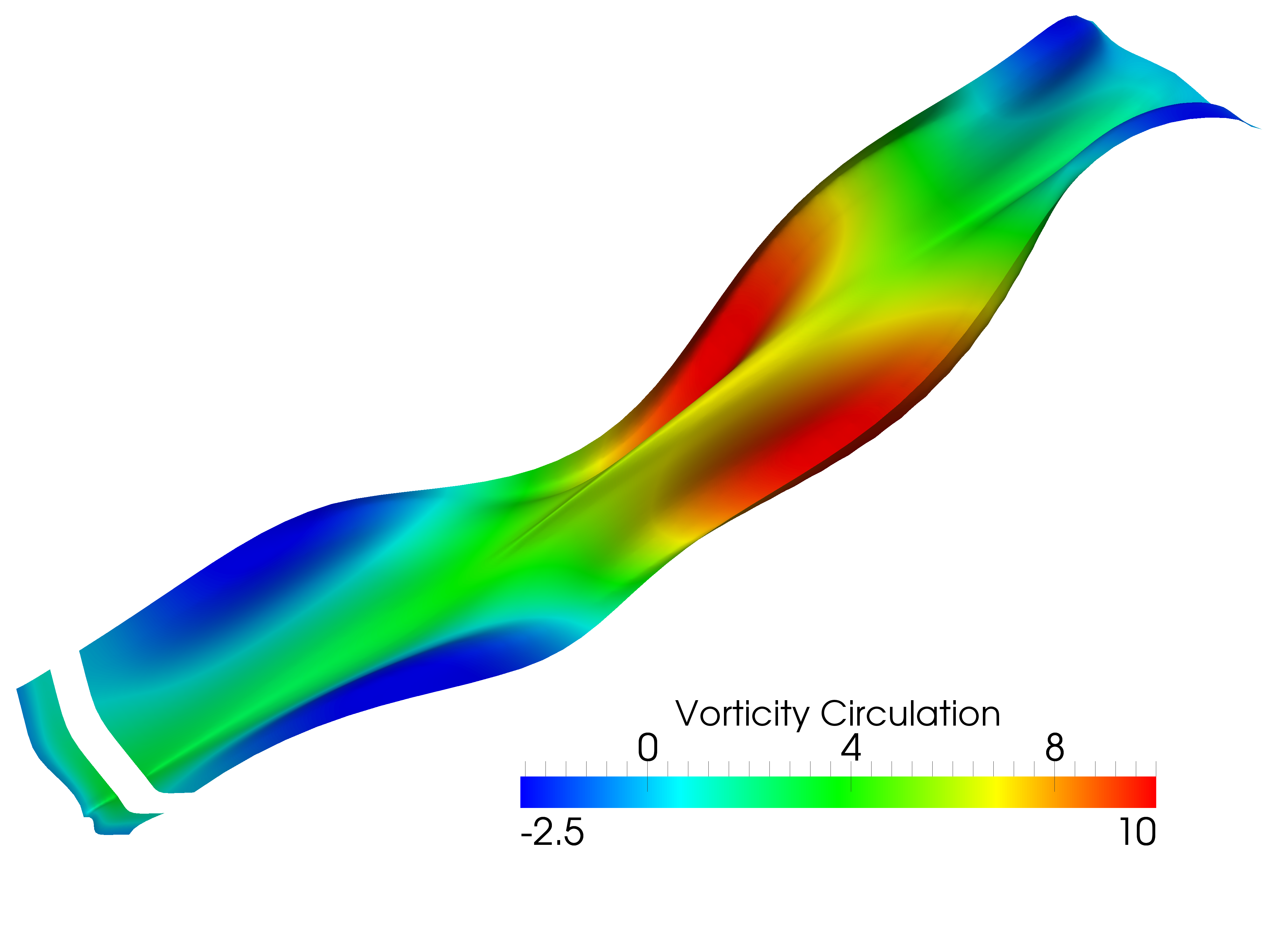}}
  \end{center}
  \caption{Wake patterns of the flapping wing obtained for the baseline and optimal shapes. Contour color levels denote the vorticity strength of the wing
  and the wake. The wake has been moved back for the sake of differentiating it from the wing.}
  \label{Figwake1}
\end{figure}

The vorticity in the wake was generated on and shed from the wing at an earlier time. As such, the wake is usually referred as the "historian" of the flow. Thus, examining the wake pattern and vorticity distribution is helpful to gain insight into the reasons why the obtained optimized shapes produce efficient flapping flights. The vorticity circulation strength of the wakes obtained for the baseline and optimal wing shapes is shown in Figure~\ref{Figwake1}. Clearly, the overall strength of the wake has increased in comparison with the baseline case, as the average aerodynamic power has increased as shown in Table \ref{OPTk01}. In particular, stronger pockets of high circulation are observed in the wake aft of the optimal shape at the middle of the down and upstrokes (i.e., $\phi \approx 0^{\circ}$). Furthermore, the vortex tip swirl is more pronounced for the baseline shape. So, the optimal shape managed to reduce the tip vortex effect which then produces higher thrust. 

\begin{figure}[ht]
  \begin{center}
      \subfigure[Baseline case]{\includegraphics[width=0.9\textwidth]{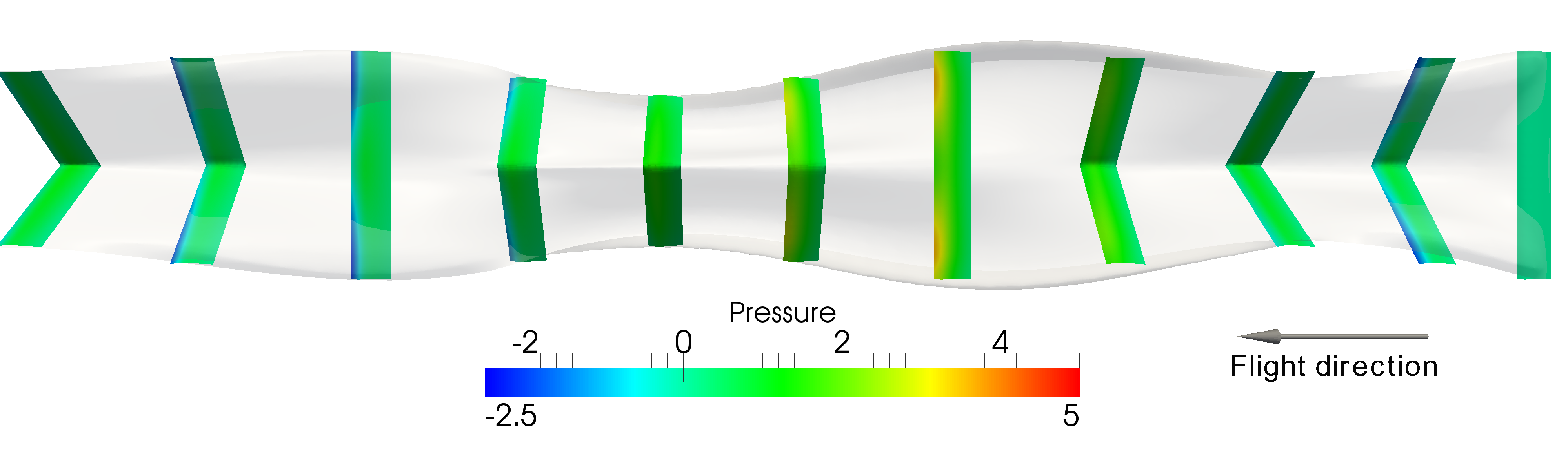}} \\
      \subfigure[Optimal case]{\includegraphics[width=0.9\textwidth]{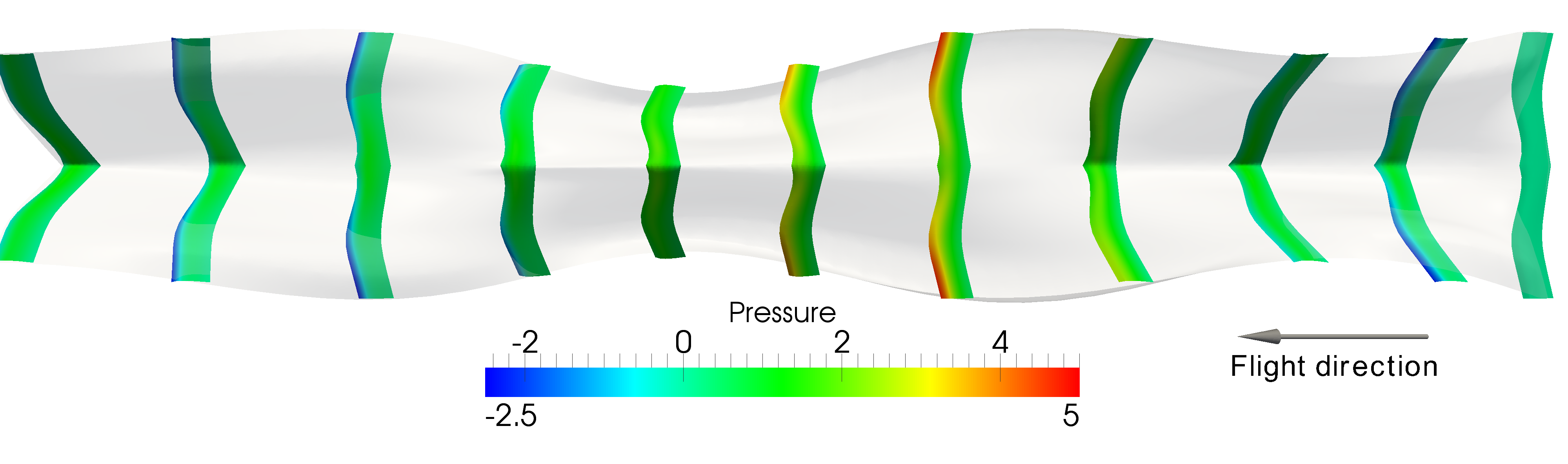}}
  \end{center}
  \caption{Pressure difference distribution over the wing surface during a flapping cycle. Top view of the flapping flight is shown.}
  \label{Figpress}
\end{figure}

The unsteady pressure difference between the lower and upper surfaces of the wing for the baseline and optimal cases is plotted in Figure~\ref{Figpress}. The pressure difference varies over the flapping cycle, giving rise to a positive pressure jump during the downstroke and negative pressure jump during the upstroke. Clearly, the highest pressure levels are obtained during the downstroke. This explains the generation of the aerodynamic loads along the two phases of the
flapping cycle as shown in Figure~\ref{Fig4}. Furthermore, higher pressure levels are achieved for the optimal case as reflective of the increase in the required aerodynamic power. For both cases (baseline and optimal), we observe a pressure peak along the leading edge. This pressure peak may lead to higher adverse pressure gradient and may be indicative of an eventual flow separation \cite{Persson2012}. This peak is mostly related to the fact that the aerodynamic model (based on potential flow) enforces the flow to be attached around the sharp leading edge of the wing.

\paragraph{Separate leading and trailing edges curvature effects}
The above results indicate that one needs to design properly the trailing edge also since the wing may benefit from the interaction with the near wake and then improve its aerodynamic performance. In an attempt to separate the individual effects of the trailing and leading edges on the wing performance, we kept one of them to be fixed at 90$^{\circ}$ angle with respect to the flight direction while allowing the other to be curved and turned on the optimizer. Again, the number of degrees of freedom is held constant at 12 and the cubic approximation is used in the B-spline representation. Only increases of 17\% and 1\% in the propulsive efficiency were obtained for the optimal shapes when curving the leading and trailing edges, separately. Clearly, both of them need to be varied simultaneously to enable superior design performances.

\paragraph{Power constraint effect}
The optimal shapes enable a significant improvement in terms of propulsive efficiency, nevertheless, they require much more aerodynamic power to perform their forward flight. Thus, to keep the same level of aerodynamic power as required for the baseline case, a power constraint is added and the optimization is reformulated as follows:
\begin{equation}
\begin{alignedat}{2}
&\text{Maximize} \quad & &\eta(\bx) \\
&\text{Subject to} \quad & &\overline{L^*}(\bx) \geq \overline{L^*}_\mathrm{bl}, \\
&&&\overline{T^*}(\bx) \geq \overline{T^*}_\mathrm{bl}, \\
&&&A(\bx) \leq A_\mathrm{bl},  \\
&&&\overline{P^*}(\bx) \leq \overline{P^*}_\mathrm{bl},  \\
&&&\max_{i,j} \abs{ \theta_{i,j}(\bx) -\tfrac{\pi}{2}}\leq \theta_\mathrm{cr}, \\
&&&\bx \in X,
\end{alignedat}
\end{equation}

Results for the efficiency $\eta$, lift $\overline{L^*}$, thrust $\overline{T^*}$, aerodynamic power $\overline{P^*}$, and area ratio $A/A_\mathrm{bl}$ obtained from the optimization case when imposing a power constraint are presented in Table~\ref{OPTP}. A lower thrust has been obtained in comparison with the power unconstrained optimization case while maintaining almost the same levels of efficiency and lift. The decrease in the power consumption with respect to previous optimization case came at the price of a loss in the thrust force.

\begin{table}[ht!]
\caption{Baseline vs. optimal results, $\kappa=0.1$. Lift, thrust, area, and power constraints are imposed. Cubic approximation is used in the B-spline representation.}\label{OPTP}
\begin{center}
\begin{tabular}{l l l l l l}
  \hline
  \hline
  Wing shape & $\eta$ & $\overline{L^*}$ & $\overline{T^*}$ & $\overline{P^*}$ & $A/A_\mathrm{bl}$\\
  \hline
  \hline
  Baseline shape & 0.219 & 4.484 & 0.185 & 0.844 & 1\\
  Optimal shape & 0.438 & 5.470 & 0.369 & 0.844 & 0.82\\
  \hline
  \hline
\end{tabular}
\end{center}
\end{table}

\begin{figure}[ht]
\begin{center}
\includegraphics[width=0.75\textwidth]{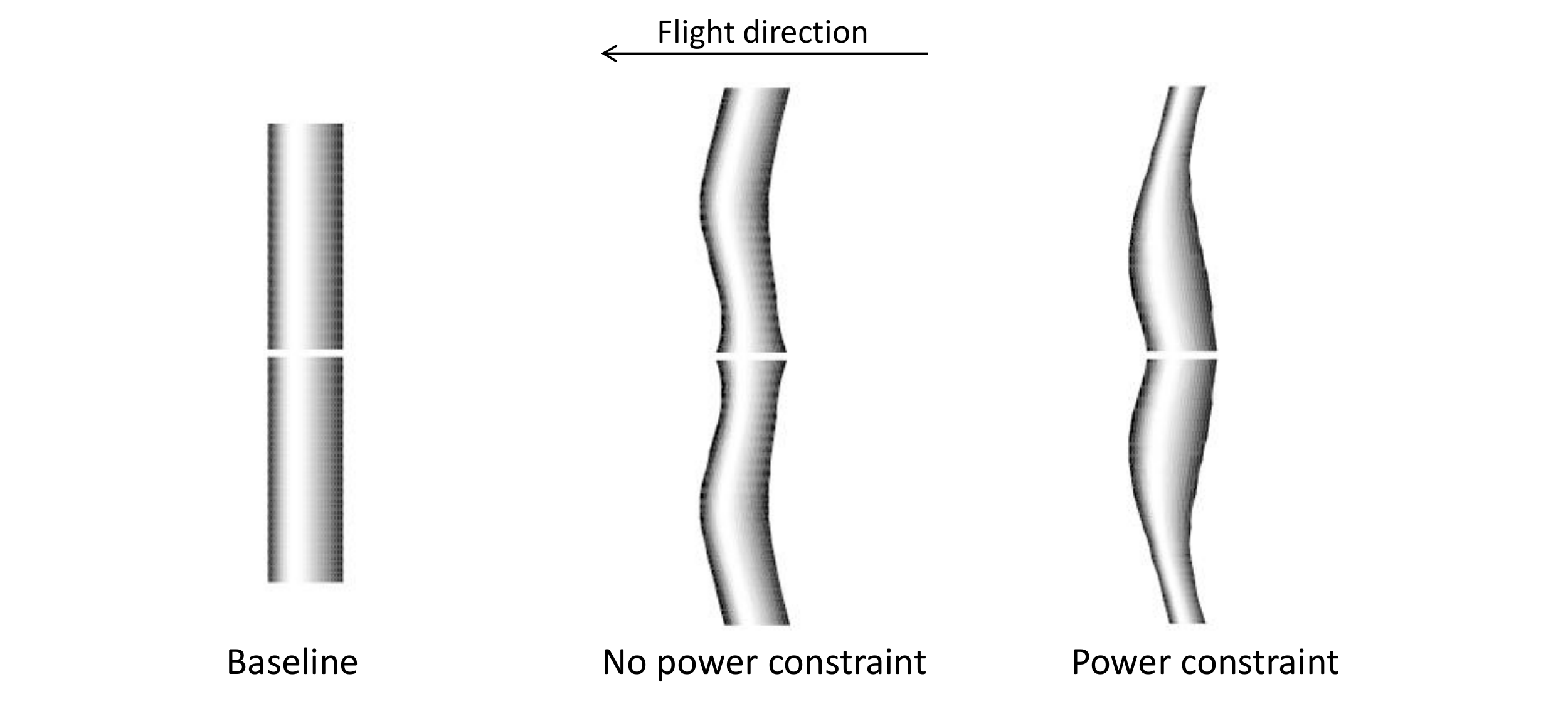}
\end{center}
\caption{Optimal wing shapes (with power constraint). Coordinates of the control points parameterizing the optimal shapes are reported in Table~\ref{Tabs8}.} \label{Fig_shape4}
\end{figure}

To satisfy the power constraint, less area is distributed to the outer part of the optimal wing as shown in Figure~\ref{Fig_shape4} while the geometric parameters defining the optimal shapes are listed in Table~\ref{Tabs8}. In particular, the wing is significantly shrunk near the tip. Since the acceleration resulting from the flapping motion is proportional to the distance from the wing root, the outer part of the wing (closer to the tip) undergoes greater accelerations and then contributes relatively more to the aerodynamic power requirement. Thus, to enable reduction in power consumption (with respect to the previous optimal configuration) while keeping the flapping flight as efficient as possible, the proportion of the wing near the tip becomes narrower. The wing's aspect ratio has increased in this power constrained optimal configuration as it did in the previous case, where the limiting factors were the wing area and the allowable range of variation adopted for the control points.

\begin{table}
\centering
\caption{Control points for the optimal shape: fixed number of degrees of freedom case and cubic approximation is used in the B-splines representation ($\kappa =0.1$). Lift, thrust, area, and power constraints are imposed.}\label{Tabs8}
{\footnotesize
\begin{tabular}{ccccccccccc}
\hline\hline
 & & \multicolumn{3}{c}{Control Points}  & & & & \multicolumn{3}{c}{Control Points}\\
$i$ & $j$ & $x$ & $y$ & $z$ & & $i$ & $j$ & $x$ & $y$ & $z$ \\
\hline\hline
 0 & 0 & 0.085173 & 0.000000 & 0.000000 &  & 3 & 0 & -0.091147 & 2.000000 & 0.000000\\
 0 & 1 & 0.237644 & 0.000000 & 0.089250 &  & 3 & 1 & 0.031862 & 2.000000 & 0.089250\\
 0 & 2 & 0.390115 & 0.000000 & 0.126467 &  & 3 & 2 & 0.154870 & 2.000000 & 0.126467\\
 0 & 3 & 0.542586 & 0.000000 & 0.034380 &  & 3 & 3 & 0.277879 & 2.000000 & 0.034380\\
 0 & 4 & 0.695058 & 0.000000 & 0.091453 &  & 3 & 4 & 0.400887 & 2.000000 & 0.091453\\
 0 & 5 & 0.847529 & 0.000000 & 0.031200 &  & 3 & 5 & 0.523895 & 2.000000 & 0.031200\\
 0 & 6 & 1.000000 & 0.000000 & 0.000000 &  & 3 & 6 & 0.646904 & 2.000000 & 0.000000\\
 1 & 0 & 0.006480 & 0.333333 & 0.000000 &  & 4 & 0 & 0.183726 & 2.666670 & 0.000000\\
 1 & 1 & 0.165743 & 0.333333 & 0.089250 &  & 4 & 1 & 0.254317 & 2.666670 & 0.089250\\
 1 & 2 & 0.325006 & 0.333333 & 0.126467 &  & 4 & 2 & 0.324909 & 2.666670 & 0.126467\\
 1 & 3 & 0.484269 & 0.333333 & 0.034380 &  & 4 & 3 & 0.395501 & 2.666670 & 0.034380\\
 1 & 4 & 0.643532 & 0.333333 & 0.091453 &  & 4 & 4 & 0.466093 & 2.666670 & 0.091453\\
 1 & 5 & 0.802795 & 0.333333 & 0.031200 &  & 4 & 5 & 0.536685 & 2.666670 & 0.031200\\
 1 & 6 & 0.962058 & 0.333333 & 0.000000 &  & 4 & 6 & 0.607276 & 2.666670 & 0.000000\\
 2 & 0 & -0.233239 & 1.000000 & 0.000000 &  & 5 & 0 & 0.383476 & 3.500000 & 0.000000\\
 2 & 1 & -0.042933 & 1.000000 & 0.089250 &  & 5 & 1 & 0.460540 & 3.500000 & 0.089250\\
 2 & 2 & 0.147373 & 1.000000 & 0.126467 &  & 5 & 2 & 0.537604 & 3.500000 & 0.126467\\
 2 & 3 & 0.337678 & 1.000000 & 0.034380 &  & 5 & 3 & 0.614668 & 3.500000 & 0.034380\\
 2 & 4 & 0.527984 & 1.000000 & 0.091453 &  & 5 & 4 & 0.691732 & 3.500000 & 0.091453\\
 2 & 5 & 0.718290 & 1.000000 & 0.031200 &  & 5 & 5 & 0.768796 & 3.500000 & 0.031200\\
 2 & 6 & 0.908595 & 1.000000 & 0.000000 &  & 5 & 6 & 0.845860 & 3.500000 & 0.000000\\
\hline\hline
\end{tabular}}
\end{table}

\begin{figure}[ht]
\begin{center}
\includegraphics[width=0.65\textwidth]{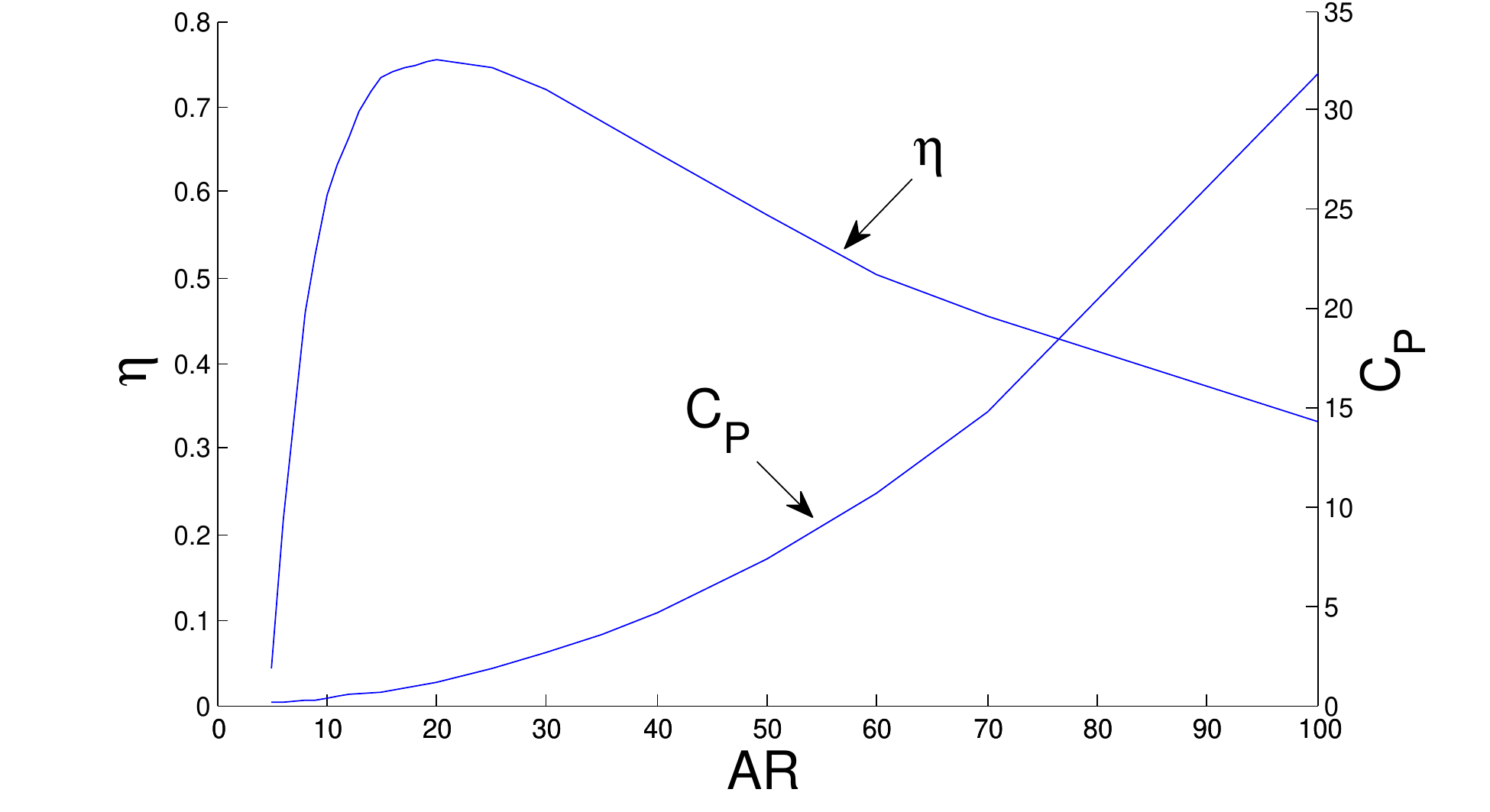}
\end{center}
\caption{Variations of the efficiency and power coefficient with the wing's aspect ratio (AR) for a rectangular wing flapping at a reduced frequency $\kappa=0.1$.} \label{Eff_AR}
\end{figure}

\paragraph{Aspect ratio effect}
To gain a better understanding of the reasons for which the obtained optimized shapes enhance the performance of the flapping flight, the impact of varying the aspect ratio is examined for a rectangular wing. Figure \ref{Eff_AR} shows the sensitivity of the efficiency and power coefficient, defined as $C_P = \frac{2P}{\rho b c U_{\infty}^3} $, to the wing's aspect ratio over a wide range of variation. We observe that the efficiency increases significantly with the aspect ratio to achieve a maximum near a value of 20, while the aerodynamic power increases monotonically when the wing's aspect ratio is increased. These results show that high-aspect-ratio wings generate more thrust and are more efficient than low-aspect-ratio wings however their flapping motion requires more aerodynamic power. This explains the identification of longer wings in the optimized configurations (the increase in the aspect ratio is limited by the area constraint and the allowable range of variation adopted for the control points) and because of the power constraint, the proportion of area in the outer region the wing has been reduced. This outer area is subjected to higher angular accelerations.

 Wings with higher aspect ratio enable better aerodynamic performance (when flying at moderate frequencies), as can be concluded from the present optimization study, however, such wings undergo greater deflections. This requires stricter structural design specifications to avoid undesirable aeroelastic effects, such as occurrence of flutter at low flight speeds. Besides, long wings have higher moment of inertia to overcome and create more parasitic drag \cite{Dommasch1961} which degrades the flight performance. All of the aforementioned reasons, which are not accounted for in our analysis, limit the consideration of high-aspect ratio wings when designing flying vehicles.

\subsubsection{High reduced-frequency case}
\begin{figure}[ht]
\begin{center}
\includegraphics[width=0.75\textwidth]{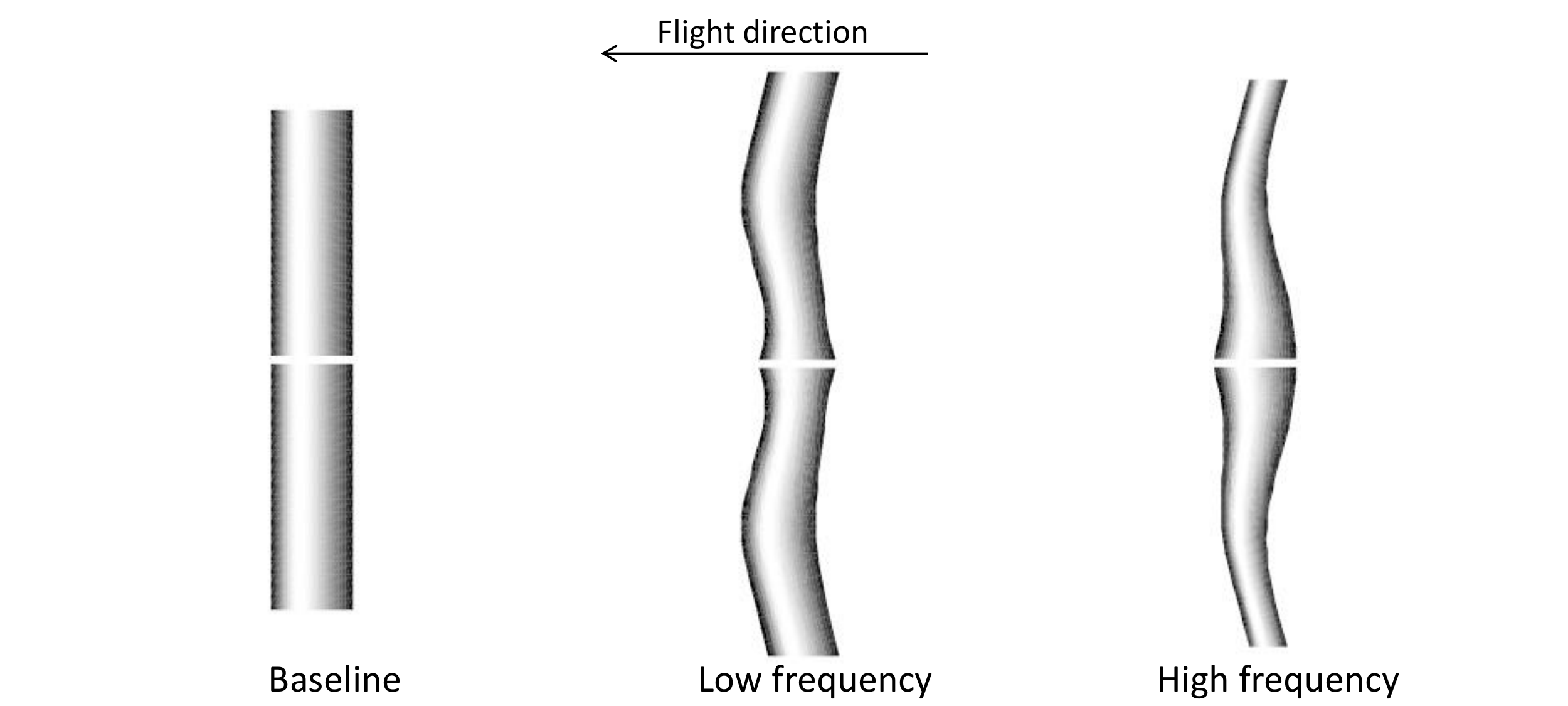}
\end{center}
\caption{Effect of reduced frequency on optimal wing shapes. Coordinates of the control points parameterizing the optimal shapes are reported in Table \ref{Tabs9}.} \label{Fig_shape5}
\end{figure}

We consider a higher reduced frequency ($\kappa=0.4$). Again, an aerodynamic mesh of 24 chordwise elements and 20 spanwise elements and 120 time steps per flapping cycle are used. This configuration is found appropriate to achieve UVLM solution independency of wing and time discretization. The previous case ($\kappa=0.1$) represents a quasi-steady flight condition, whereas in the current case unsteady effects are much more pronounced. The objective is to examine the impact of the reduced frequency on the optimal shapes. As expected, flying at higher reduced frequency involves more aerodynamic power. In fact, increasing the reduced frequency $\kappa$ leads to higher angular acceleration
for the wing and consequently for the surrounding fluid (noncirculatory), which would require additional power to accelerate it. This noncirculatory fluid is pushed by the wing and leads to what is referred as the added mass effects. Looking at the optimal shapes obtained for the two reduced frequencies given in Figure~\ref{Fig_shape5}, we observe that the high-frequency wing shape is more tilted close the wing tip, resulting in a net drop in the area (as seen in Table~\ref{OPTk04}). The geometric parameters controlling the optimal shapes are listed in Table~\ref{Tabs9}. The optimal values of the lift, thrust, aerodynamic power, and area ratio are reported in Table~\ref{OPTk04}. A moderate improvement in terms of propulsive efficiency ($\sim 28 \%$ increase) is obtained. Optimal results show different trends with respect to the low-frequency case. In fact, the area constraint is not active anymore. The thrust constraint is active and a lower aerodynamic power are obtained such that propulsive efficiency improves (from 0.457 to 0.576). Unlike the low-frequency case, the improvement in the propulsive efficiency is entirely due to a decrease in the input aerodynamic power.

\begin{table}
\centering
\caption{Control points for the optimal shape: fixed number of degrees of freedom case and cubic approximation is used in the B-splines representation ($\kappa =0.4$). Lift, thrust, and area constraints are imposed.}\label{Tabs9}
{\footnotesize
\begin{tabular}{ccccccccccc}
\hline\hline
 & & \multicolumn{3}{c}{Control Points}  & & & & \multicolumn{3}{c}{Control Points}\\
$i$ & $j$ & $x$ & $y$ & $z$ & & $i$ & $j$ & $x$ & $y$ & $z$ \\
\hline\hline
 0 & 0 & 0.000227 & 0.000000 & 0.000000 &  & 3 & 0 & 0.030328 & 2.000000 & 0.000000\\
 0 & 1 & 0.166856 & 0.000000 & 0.089250 &  & 3 & 1 & 0.120033 & 2.000000 & 0.089250\\
 0 & 2 & 0.333484 & 0.000000 & 0.126467 &  & 3 & 2 & 0.209738 & 2.000000 & 0.126467\\
 0 & 3 & 0.500113 & 0.000000 & 0.034380 &  & 3 & 3 & 0.299443 & 2.000000 & 0.034380\\
 0 & 4 & 0.666742 & 0.000000 & 0.091453 &  & 3 & 4 & 0.389149 & 2.000000 & 0.091453\\
 0 & 5 & 0.833371 & 0.000000 & 0.031200 &  & 3 & 5 & 0.478854 & 2.000000 & 0.031200\\
 0 & 6 & 1.000000 & 0.000000 & 0.000000 &  & 3 & 6 & 0.568559 & 2.000000 & 0.000000\\
 1 & 0 & 0.107023 & 0.333333 & 0.000000 &  & 4 & 0 & 0.256388 & 2.666670 & 0.000000\\
 1 & 1 & 0.258868 & 0.333333 & 0.089250 &  & 4 & 1 & 0.329331 & 2.666670 & 0.089250\\
 1 & 2 & 0.410713 & 0.333333 & 0.126467 &  & 4 & 2 & 0.402274 & 2.666670 & 0.126467\\
 1 & 3 & 0.562558 & 0.333333 & 0.034380 &  & 4 & 3 & 0.475218 & 2.666670 & 0.034380\\
 1 & 4 & 0.714404 & 0.333333 & 0.091453 &  & 4 & 4 & 0.548161 & 2.666670 & 0.091453\\
 1 & 5 & 0.866249 & 0.333333 & 0.031200 &  & 4 & 5 & 0.621104 & 2.666670 & 0.031200\\
 1 & 6 & 1.018090 & 0.333333 & 0.000000 &  & 4 & 6 & 0.694047 & 2.666670 & 0.000000\\
 2 & 0 & 0.145300 & 1.000000 & 0.000000 &  & 5 & 0 & 0.433250 & 3.398740 & 0.000000\\
 2 & 1 & 0.262970 & 1.000000 & 0.089250 &  & 5 & 1 & 0.510371 & 3.398740 & 0.089250\\
 2 & 2 & 0.380640 & 1.000000 & 0.126467 &  & 5 & 2 & 0.587493 & 3.398740 & 0.126467\\
 2 & 3 & 0.498310 & 1.000000 & 0.034380 &  & 5 & 3 & 0.664615 & 3.398740 & 0.034380\\
 2 & 4 & 0.615980 & 1.000000 & 0.091453 &  & 5 & 4 & 0.741737 & 3.398740 & 0.091453\\
 2 & 5 & 0.733650 & 1.000000 & 0.031200 &  & 5 & 5 & 0.818858 & 3.398740 & 0.031200\\
 2 & 6 & 0.851320 & 1.000000 & 0.000000 &  & 5 & 6 & 0.895980 & 3.398740 & 0.000000\\
\hline\hline
\end{tabular}}
\end{table}

\begin{table}[ht!]
\caption{Baseline vs. optimal results, $\kappa=0.4$. Lift, thrust, and area constraints are imposed. Cubic approximation is used in the B-spline representation.}\label{OPTk04}
\begin{center}
\begin{tabular}{l l l l l l}
  \hline
  \hline
  Wing shape & $\eta$ & $\overline{L^*}$ & $\overline{T^*}$ & $\overline{P^*}$ & $A/A_\mathrm{bl}$\\
  \hline
  \hline
  Baseline shape & 0.457 & 6.227 & 4.108 & 8.98 & 1\\
  Optimal shape & 0.576 & 7.294 & 4.108 & 7.13 & 0.716\\
  \hline
  \hline
\end{tabular}
\end{center}
\end{table}

\begin{figure}
  \begin{center}
  \begin{center}
      \subfigure[Lift]{\includegraphics[width=0.45\textwidth]{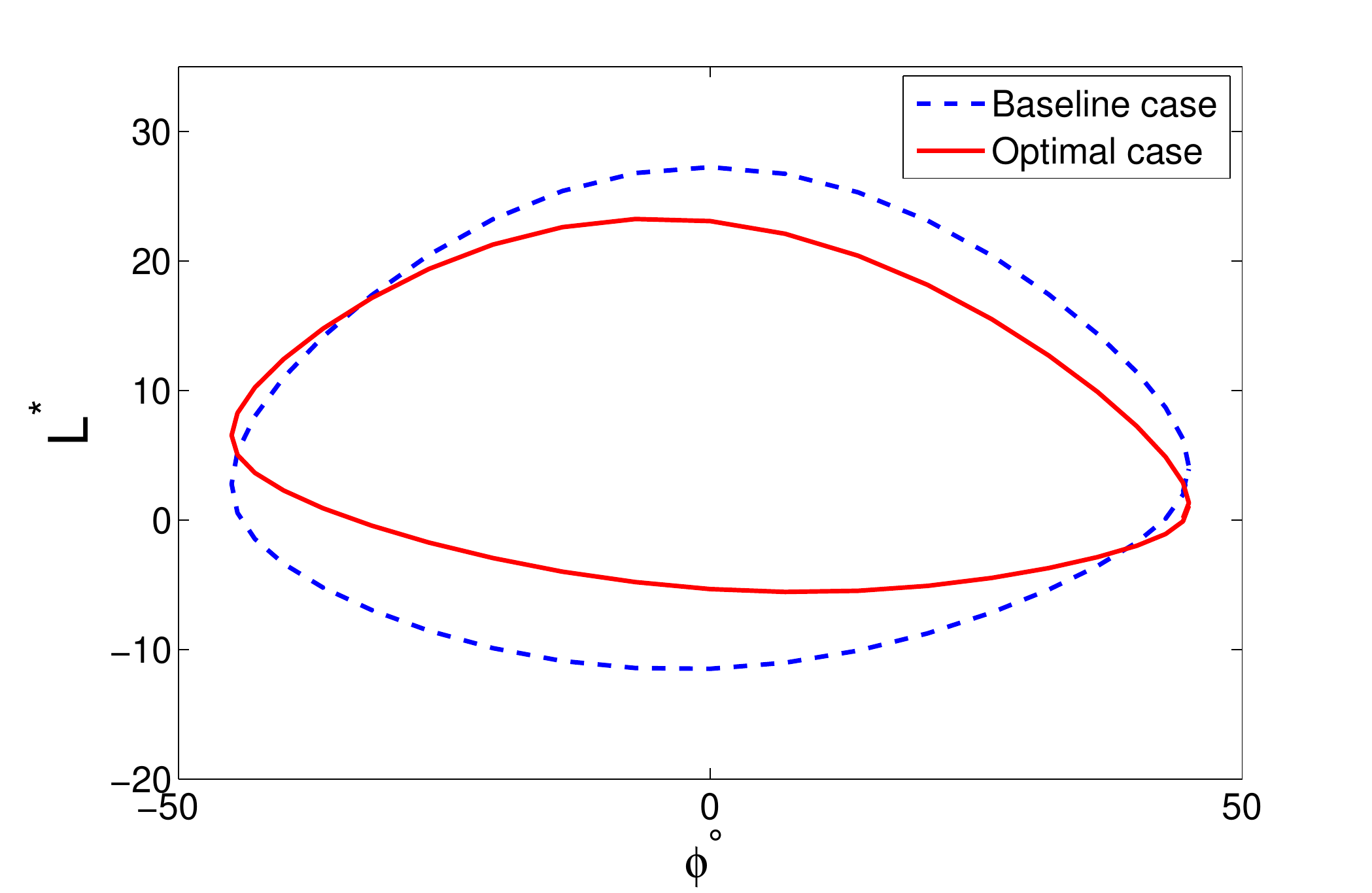}\label{Fig5a}}
      \subfigure[Thrust]{\includegraphics[width=0.45\textwidth]{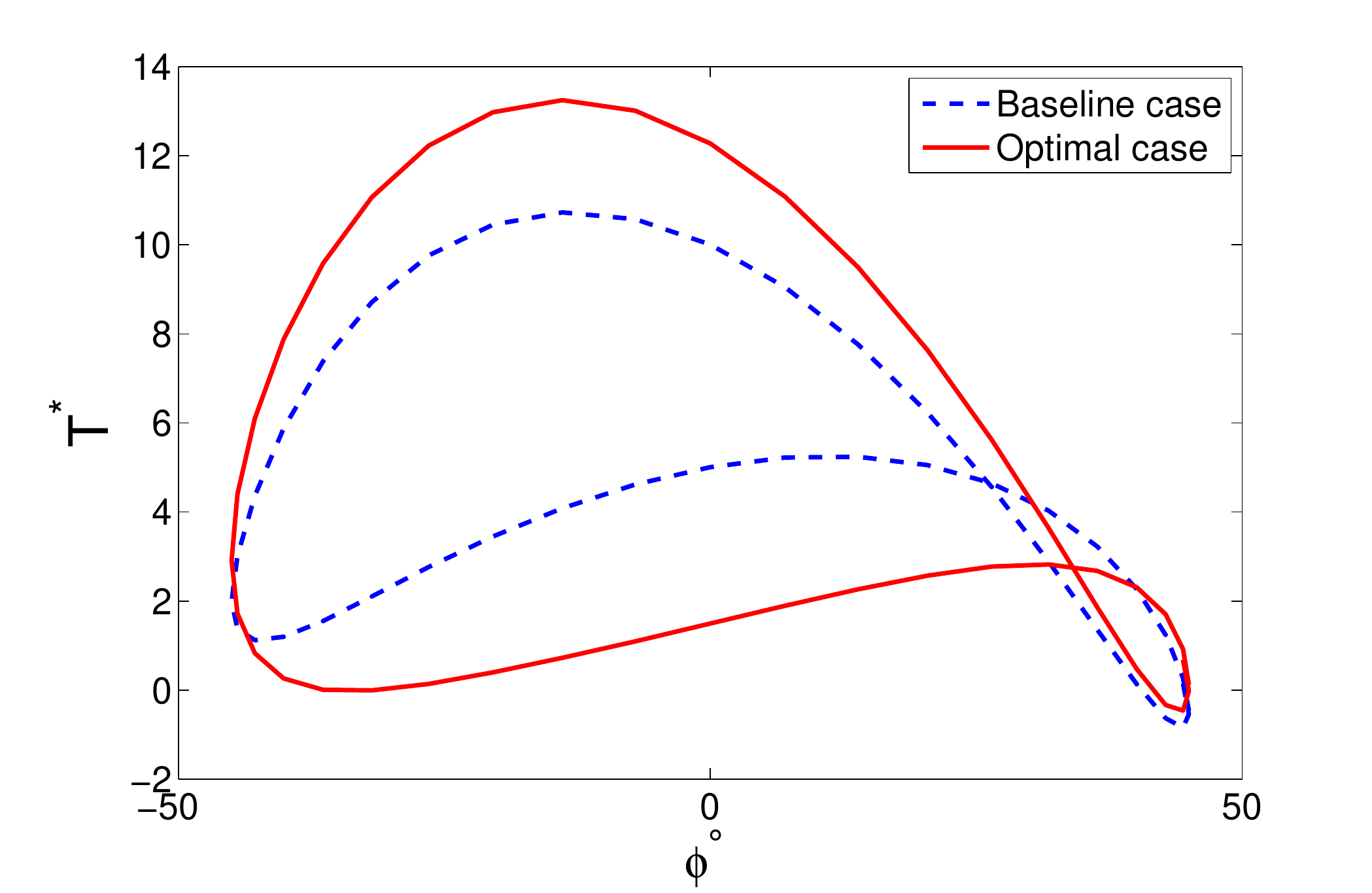}\label{Fig5b}}
      \subfigure[Power]{\includegraphics[width=0.45\textwidth]{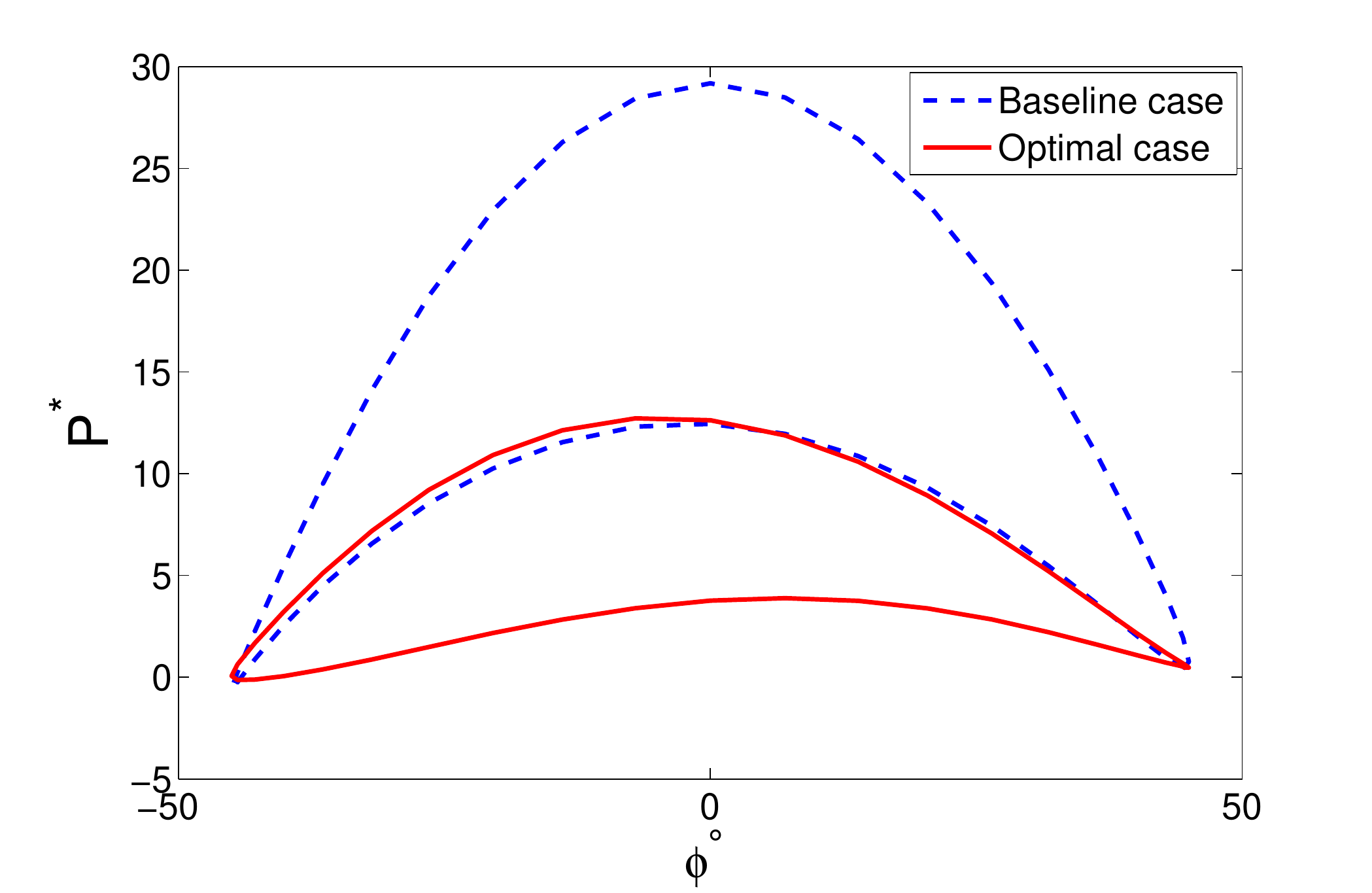}\label{Fig5c}}
  \end{center}
  \end{center}
  \caption{Lift, thrust, and power plotted versus the flapping angle $\phi$ for the baseline (rectangular) and optimal shapes ($\kappa=0.4$).}
  \label{Fig5}
\end{figure}

The lift, thrust, and power phase plots obtained at $\kappa=0.4$ are shown in Figure~\ref{Fig5}. Again, the majority of the positive lift is generated during the downstroke, however now it reaches negative values during the upstroke. This is caused by the fact that the contribution of the flapping motion to the effective angle of attack is larger than that of the fixed pitch angle at the wing root, resulting in negative angles during the upstroke.

\begin{figure}[ht]
  \centering
  \subfigure[Baseline case]{\includegraphics[width=0.48\textwidth]{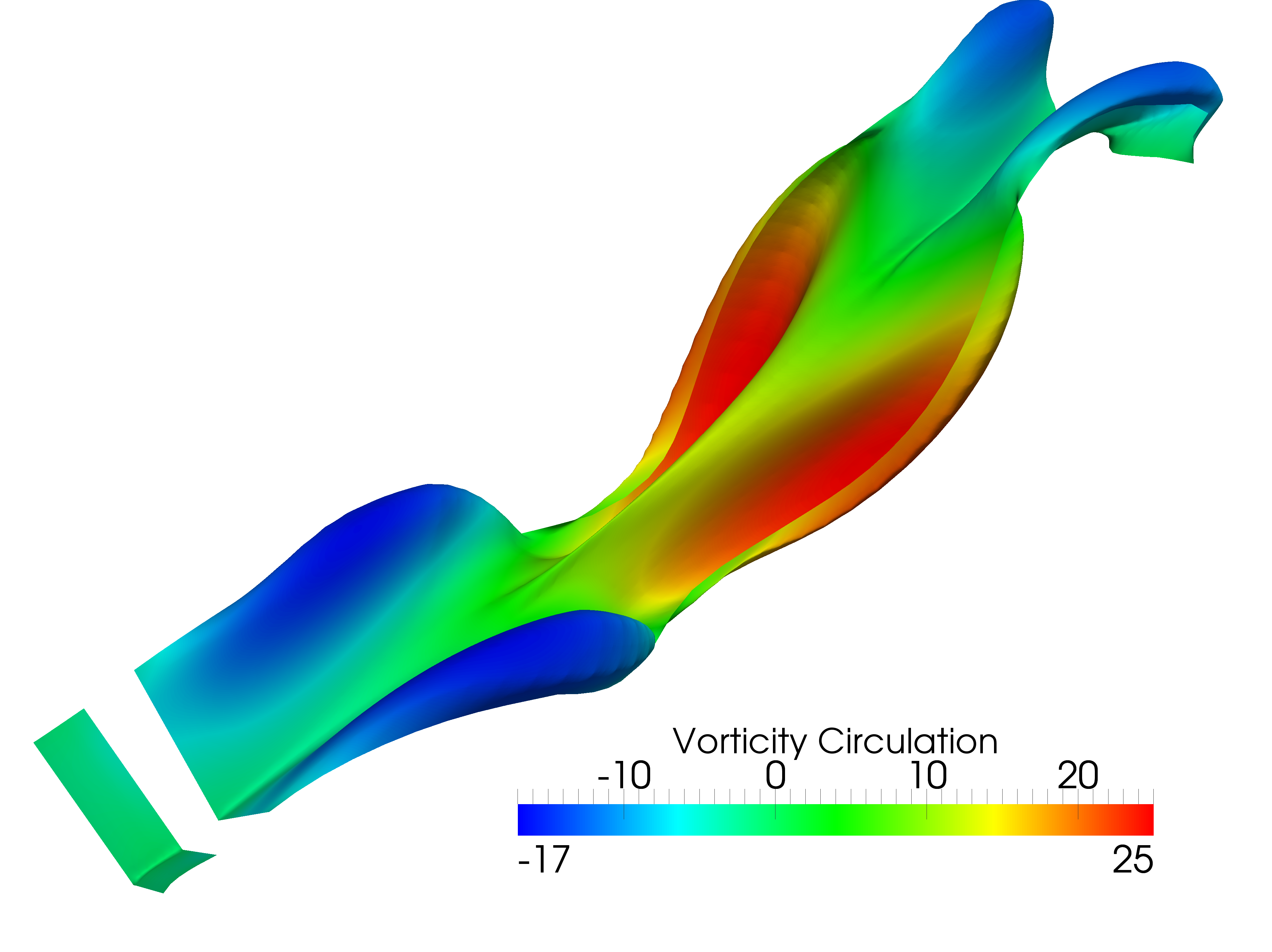}}
  \subfigure[Optimal case]{\includegraphics[width=0.48\textwidth]{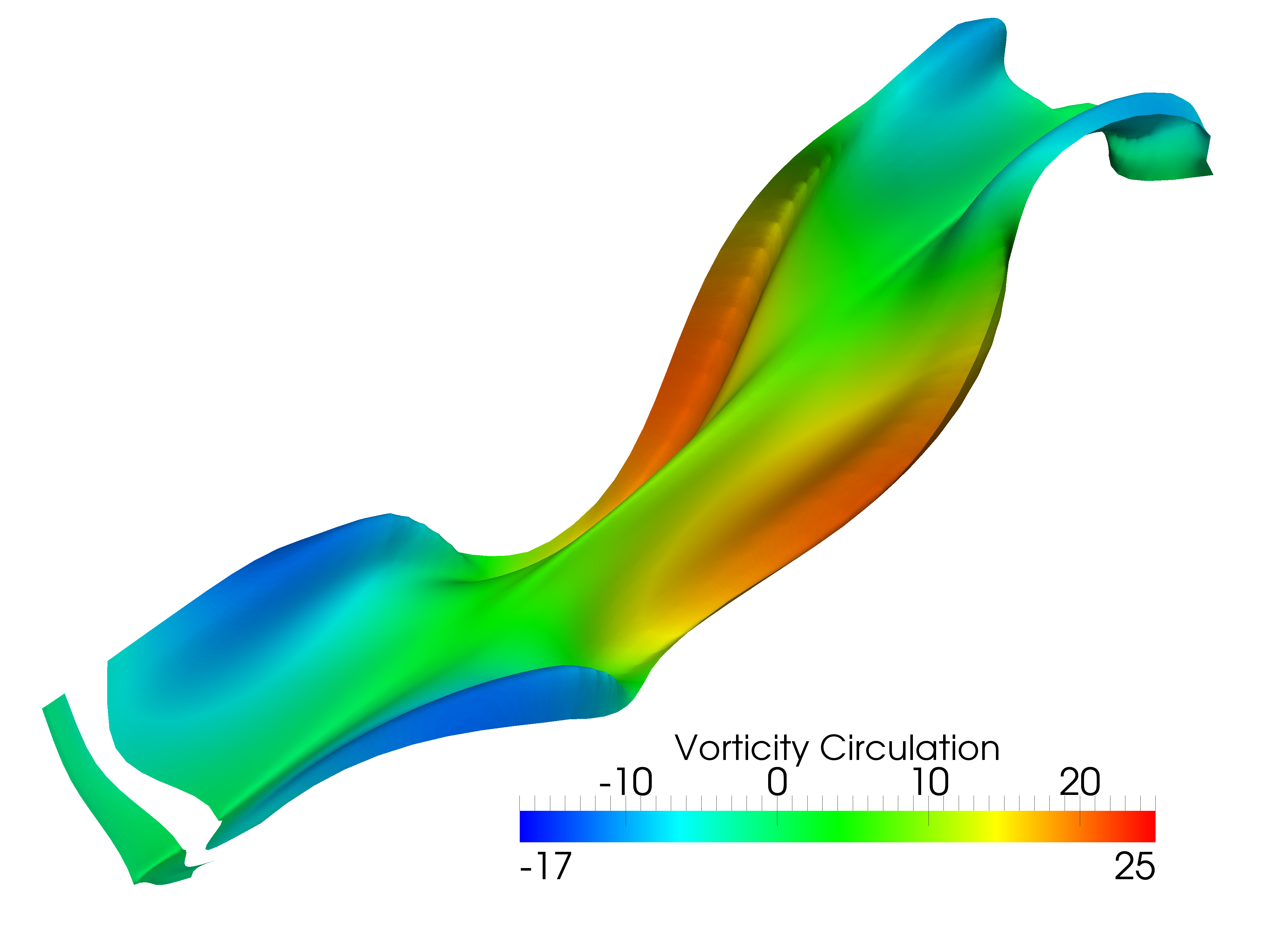}}
  \caption{Wake patterns of the flapping wing obtained for the baseline and optimal shapes ($\kappa=0.4$). Contour color levels denote the vorticity strength of the wing
  and the wake. The wake has been moved back for the sake of differentiating it from the wing.}
  \label{Figwake2}
\end{figure}

The wakes trailing the baseline and optimal
wing configurations obtained for high frequency ($\kappa=0.4$) are shown in Figure~\ref{Figwake2} where the circulation strength of those wakes can be observed.
The vortex rings in the wake developed over a flapping cycle are located very close to the wing. Then, the wing-wake interactions have more significant effects on the aerodynamic behavior of the flapping wing. In comparison with the baseline case, the vortex rings in the wake aft of the optimal flapping wing are weak as reflective of the drop in required input power.

\subsection{Camber effect}
We allow the camber line to vary by introducing perturbations to the interior control points shown in Figure \ref{Skematic} along the z-direction and turn on the optimizer. The camber line is kept the same at all cross sections along the spanwise direction. Moderate variations are considered to avoid wing cross sections with high angle of attack (with respect to the incoming freestream) that may initiate flow separation which cannot be captured in the present aerodynamic model. Again, the use of B-spline representation enables to smoothly discretize wing camber line with few degrees of freedom and then a variety of camber lines can be obtained rather than sticking to NACA profiles. The optimal results for the efficiency and aerodynamic loads and power are summarized in Tables \ref{OPTcamber} and \ref{OPTcamber2} for $\kappa=0.1$ and $\kappa=0.4$, respectively. As expected, the camber greatly affects the aerodynamic characteristics such as power and thrust. At low frequency, significantly higher thrust and efficiency can be obtained by optimizing the camber shape. The average value of the lift force in the optimized shape, obtained when introducing variations to the camber line, remained equal to the value in the baseline case. In fact, the optimizer would decrease the lift if it could, because this constraint is always active. On the other hand, higher thrust is achieved in the optimized configuration. So, there is a natural trade-off between lift and efficiency, but not so for thrust and efficiency. At high frequency, the optimization drives the design to a shape configuration with slightly lower input aerodynamic power, though the time-averaged thrust remained equal to that obtained for the baseline case, yielding a mild raise in the propulsive efficiency in comparison with the previous optimal shape obtained without varying the camber line.

\begin{figure}[ht]
\begin{center}
\includegraphics[width=0.75\textwidth]{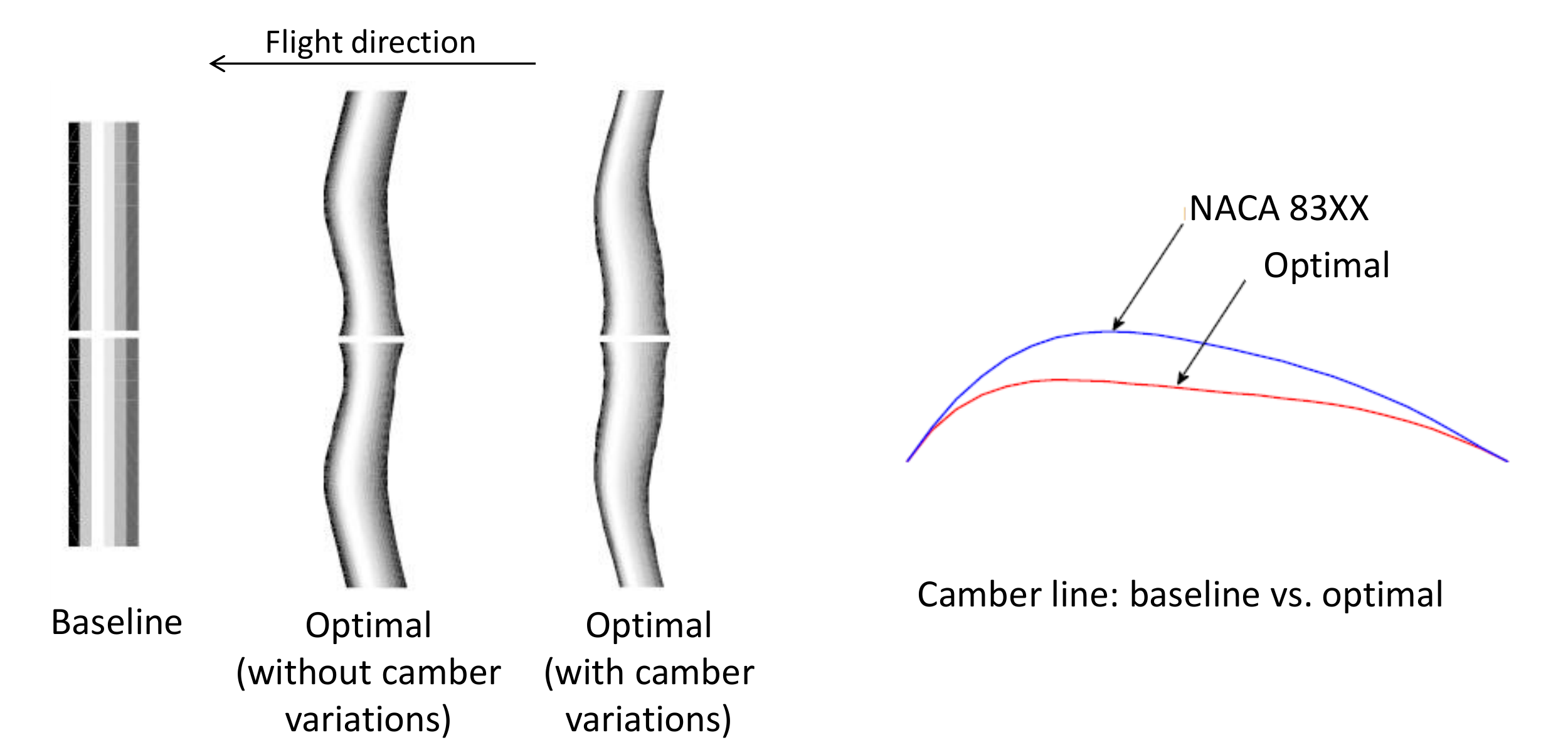}
\end{center}
\caption{Optimal wing shape and camber line ($\kappa=0.1$).  Cubic approximation is used in the B-spline representation. Coordinates of the control points parameterizing the optimal shapes are reported in Table \ref{Tabs10}.} \label{Camber}
\end{figure}

\begin{figure}[ht]
\begin{center}
\includegraphics[width=0.75\textwidth]{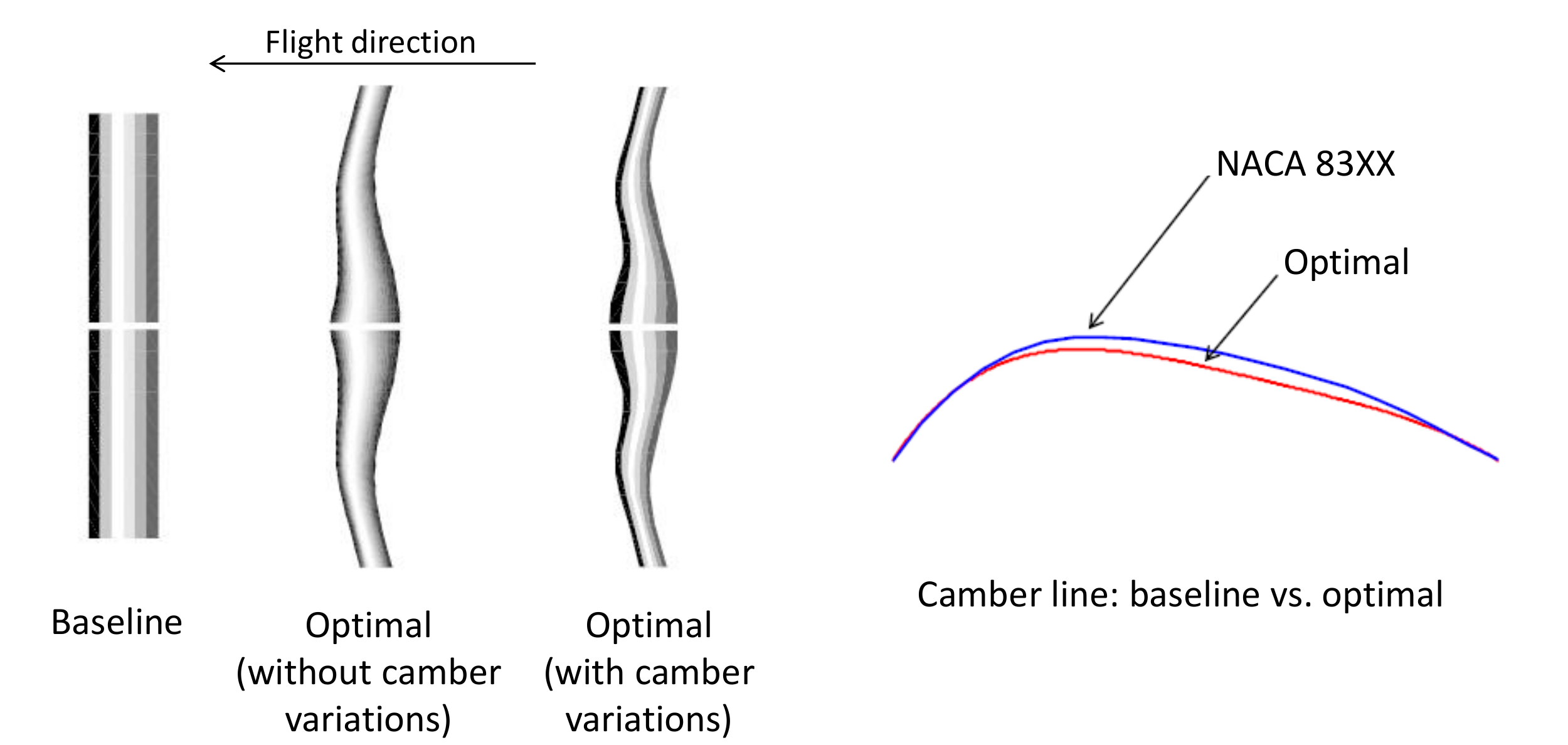}
\end{center}
\caption{Optimal wing shape and camber line ($\kappa=0.4$).  Cubic approximation is used in the B-spline representation. Coordinates of the control points parameterizing the optimal shapes are reported in Table \ref{Tabs11}.} \label{Camber2}
\end{figure}

The optimal wing shapes and their camber lines are shown in Figures \ref{Camber} and \ref{Camber2} for $\kappa=0.1$ and $\kappa=0.4$, respectively, while the geometric configurations are detailed in Tables~\ref{Tabs10} and~\ref{Tabs11}. The optimal shapes show qualitative similarity along the spanwise direction with respect to the previous optimal shape obtained without changing the camber line. The optimal camber line is thinner, in particular for the low frequency case, than the NACA 83XX profile resulting in lower effective angle of attack. The change in the camber line enabled better performance in terms of efficiency but induced a reduction in lift generation (although the lift is higher than its value obtained for a rectangular wing). These results provide guidance on how to select the wing planform and the camber based on the nature of the mission and its aerodynamic requirements.

\begin{table}[ht!]
\caption{Baseline vs. optimal results, $\kappa=0.1$. Lift, thrust, and area constraints are imposed. Cubic approximation is used in the B-spline representation.}\label{OPTcamber}
\begin{center}
\begin{tabular}{l l l l l l}
  \hline
  \hline
  Wing shape & $\eta$ & $\overline{L^*}$ & $\overline{T^*}$ & $\overline{P^*}$ & $A/A_\mathrm{bl}$\\
  \hline
  \hline
  Baseline shape & 0.219 & 4.484 & 0.185 & 0.844 & 1 \\
  Optimal shape without camber variations & 0.441 & 5.398 & 0.506 & 1.146 & 1 \\
  \rowcolor[gray]{.9} Optimal shape with camber variations & 0.533 & 4.484 & 0.526 & 0.986 & 0.93 \\
  \hline
  \hline
\end{tabular}
\end{center}
\end{table}

\begin{table}[ht!]
\caption{Baseline vs. optimal results, $\kappa=0.4$. Lift, thrust, and area constraints are imposed. Cubic approximation is used in the B-spline representation.}\label{OPTcamber2}
\begin{center}
\begin{tabular}{l l l l l l}
  \hline
  \hline
  Wing shape & $\eta$ & $\overline{L^*}$ & $\overline{T^*}$ & $\overline{P^*}$ & $A/A_\mathrm{bl}$\\
  \hline
  \hline
  Baseline shape & 0.457 & 6.227 & 4.108 & 8.98 & 1\\
  Optimal shape without camber variations & 0.576 & 7.294 & 4.108 & 7.13 & 0.716\\
  \rowcolor[gray]{.9} Optimal shape with camber variations & 0.612 & 6.910 & 4.108 & 6.499 & 0.693 \\
  \hline
  \hline
\end{tabular}
\end{center}
\end{table}

\begin{table}
\centering
\caption{Control points for the optimal shape: fixed number of degrees of freedom case and cubic approximation is used in the B-splines representation ($\kappa =0.1$ and with camber variations). Lift, thrust, and area constraints are imposed.}\label{Tabs10}
{\footnotesize
\begin{tabular}{ccccccccccc}
\hline\hline
 & & \multicolumn{3}{c}{Control Points}  & & & & \multicolumn{3}{c}{Control Points}\\
$i$ & $j$ & $x$ & $y$ & $z$ & & $i$ & $j$ & $x$ & $y$ & $z$ \\
\hline\hline
 0 & 0 & 0.003370 & 0.000000 & 0.000000 &  & 3 & 0 & -0.199021 & 2.000000 & 0.000000\\
 0 & 1 & 0.169475 & 0.000000 & 0.089250 &  & 3 & 1 & -0.059942 & 2.000000 & 0.089250\\
 0 & 2 & 0.335580 & 0.000000 & 0.042377 &  & 3 & 2 & 0.079137 & 2.000000 & 0.042377\\
 0 & 3 & 0.501685 & 0.000000 & 0.035764 &  & 3 & 3 & 0.218216 & 2.000000 & 0.035764\\
 0 & 4 & 0.667790 & 0.000000 & 0.053259 &  & 3 & 4 & 0.357295 & 2.000000 & 0.053259\\
 0 & 5 & 0.833895 & 0.000000 & 0.031200 &  & 3 & 5 & 0.496374 & 2.000000 & 0.031200\\
 0 & 6 & 1.000000 & 0.000000 & 0.000000 &  & 3 & 6 & 0.635453 & 2.000000 & 0.000000\\
 1 & 0 & 0.091649 & 0.333333 & 0.000000 &  & 4 & 0 & 0.044826 & 2.666670 & 0.000000\\
 1 & 1 & 0.236577 & 0.333333 & 0.089250 &  & 4 & 1 & 0.170684 & 2.666670 & 0.089250\\
 1 & 2 & 0.381505 & 0.333333 & 0.042377 &  & 4 & 2 & 0.296542 & 2.666670 & 0.042377\\
 1 & 3 & 0.526433 & 0.333333 & 0.035764 &  & 4 & 3 & 0.422401 & 2.666670 & 0.035764\\
 1 & 4 & 0.671361 & 0.333333 & 0.053259 &  & 4 & 4 & 0.548259 & 2.666670 & 0.053259\\
 1 & 5 & 0.816289 & 0.333333 & 0.031200 &  & 4 & 5 & 0.674117 & 2.666670 & 0.031200\\
 1 & 6 & 0.961217 & 0.333333 & 0.000000 &  & 4 & 6 & 0.799975 & 2.666670 & 0.000000\\
 2 & 0 & 0.001093 & 1.000000 & 0.000000 &  & 5 & 0 & 0.262770 & 3.500000 & 0.000000\\
 2 & 1 & 0.178065 & 1.000000 & 0.089250 &  & 5 & 1 & 0.376402 & 3.500000 & 0.089250\\
 2 & 2 & 0.355037 & 1.000000 & 0.042377 &  & 5 & 2 & 0.490034 & 3.500000 & 0.042377\\
 2 & 3 & 0.532009 & 1.000000 & 0.035764 &  & 5 & 3 & 0.603666 & 3.500000 & 0.035764\\
 2 & 4 & 0.708981 & 1.000000 & 0.053259 &  & 5 & 4 & 0.717298 & 3.500000 & 0.053259\\
 2 & 5 & 0.885953 & 1.000000 & 0.031200 &  & 5 & 5 & 0.830930 & 3.500000 & 0.031200\\
 2 & 6 & 1.062930 & 1.000000 & 0.000000 &  & 5 & 6 & 0.944562 & 3.500000 & 0.000000\\
\hline\hline
\end{tabular}}
\end{table}
\begin{table}
\centering
\caption{Control points for the optimal shape: fixed number of degrees of freedom case and cubic approximation is used in the B-splines representation ($\kappa =0.4$ and with camber variations). Lift, thrust, and area constraints are imposed.}\label{Tabs11}
{\footnotesize
\begin{tabular}{ccccccccccc}
\hline\hline
 & & \multicolumn{3}{c}{Control Points}  & & & & \multicolumn{3}{c}{Control Points}\\
$i$ & $j$ & $x$ & $y$ & $z$ & & $i$ & $j$ & $x$ & $y$ & $z$ \\
\hline\hline
 0 & 0 & 0.028839 & 0.000000 & 0.000000 &  & 3 & 0 & -0.043142 & 2.000000 & 0.000000\\
 0 & 1 & 0.190699 & 0.000000 & 0.089250 &  & 3 & 1 & 0.061513 & 2.000000 & 0.089250\\
 0 & 2 & 0.352559 & 0.000000 & 0.115212 &  & 3 & 2 & 0.166168 & 2.000000 & 0.115212\\
 0 & 3 & 0.514420 & 0.000000 & 0.021558 &  & 3 & 3 & 0.270823 & 2.000000 & 0.021558\\
 0 & 4 & 0.676280 & 0.000000 & 0.072092 &  & 3 & 4 & 0.375477 & 2.000000 & 0.072092\\
 0 & 5 & 0.838140 & 0.000000 & 0.031200 &  & 3 & 5 & 0.480132 & 2.000000 & 0.031200\\
 0 & 6 & 1.000000 & 0.000000 & 0.000000 &  & 3 & 6 & 0.584787 & 2.000000 & 0.000000\\
 1 & 0 & 0.010549 & 0.333333 & 0.000000 &  & 4 & 0 & 0.285517 & 2.666670 & 0.000000\\
 1 & 1 & 0.183619 & 0.333333 & 0.089250 &  & 4 & 1 & 0.337391 & 2.666670 & 0.089250\\
 1 & 2 & 0.356689 & 0.333333 & 0.115212 &  & 4 & 2 & 0.389265 & 2.666670 & 0.115212\\
 1 & 3 & 0.529759 & 0.333333 & 0.021558 &  & 4 & 3 & 0.441138 & 2.666670 & 0.021558\\
 1 & 4 & 0.702830 & 0.333333 & 0.072092 &  & 4 & 4 & 0.493012 & 2.666670 & 0.072092\\
 1 & 5 & 0.875900 & 0.333333 & 0.031200 &  & 4 & 5 & 0.544886 & 2.666670 & 0.031200\\
 1 & 6 & 1.048970 & 0.333333 & 0.000000 &  & 4 & 6 & 0.596760 & 2.666670 & 0.000000\\
 2 & 0 & 0.401749 & 1.000000 & 0.000000 &  & 5 & 0 & 0.474763 & 3.500000 & 0.000000\\
 2 & 1 & 0.482217 & 1.000000 & 0.089250 &  & 5 & 1 & 0.537329 & 3.500000 & 0.089250\\
 2 & 2 & 0.562685 & 1.000000 & 0.115212 &  & 5 & 2 & 0.599895 & 3.500000 & 0.115212\\
 2 & 3 & 0.643152 & 1.000000 & 0.021558 &  & 5 & 3 & 0.662461 & 3.500000 & 0.021558\\
 2 & 4 & 0.723620 & 1.000000 & 0.072092 &  & 5 & 4 & 0.725027 & 3.500000 & 0.072092\\
 2 & 5 & 0.804088 & 1.000000 & 0.031200 &  & 5 & 5 & 0.787593 & 3.500000 & 0.031200\\
 2 & 6 & 0.884556 & 1.000000 & 0.000000 &  & 5 & 6 & 0.850159 & 3.500000 & 0.000000\\
\hline\hline
\end{tabular}}
\end{table}

Our study gives comparable optimal efficiencies to those obtained from previous studies \cite{Stanford2010,Ghommem2012} where the improvement of flight performance was achieved by introducing a dynamic shape deformation (usually referred as active shape morphing) which is based on imposing a prescribed bending and twisting. This shows again the importance of wing planform selection for an efficient design of flapping wing vehicles. Nevertheless, the optimal shapes reported here do not need to be morphed during forward flight to achieve their improved performance.

\section{Summary and Conclusion}
 The wing shape/planform presents a critical determinant of performance in flapping flight. In this work, we use a three-dimensional version of the unsteady vortex lattice method (UVLM) to simulate the aerodynamic response of a flapping wing in forward flight and adopt B-spline representation to model the wing geometry. First, we carry out a convergence analysis to determine the appropriate time step and aerodynamic mesh sizes which are needed to reach UVLM solution independence to discretization. In particular, selecting the proper time step was important to observe the UVLM solution when refining the aerodynamic mesh. Then, we combine UVLM with a gradient-based optimizer to identify a set of efficient shapes that optimizes the aerodynamic performance. Flow simulations using the optimal wing shapes indicate that changes in the shape have significant effects on relevant averaged quantities such as lift, thrust, and aerodynamic power. Increasing the number of variables (i.e., providing the wing shape with a greater degree of spatial freedom) enables increasingly superior designs. The optimization study shows that the camber line and the leading and trailing edges represent the key wing shape parameters that control the generation of the aerodynamic loads and flight performance depending on the nature of the mission assigned to the flying vehicle. Furthermore, the optimal shapes show significant dependency on the reduced frequency. At low frequency, the optimization pushes the design to a shape configuration with substantial increase in the time-averaged thrust, while the average aerodynamic power is increased, resulting in a significant increase in the propulsive efficiency. To keep the same level of aerodynamic power as required for the baseline rectangular wing, a power constraint is added to the optimization problem. A reduction in the outer region of the wing near the tip is observed in the optimal shape. This outer area is subjected to higher angular accelerations. At high frequency, the optimization drives the design to a shape configuration with lower input aerodynamic power, though the time-averaged thrust remained equal to that obtained for the baseline case, resulting in a moderate improvement in terms of propulsive efficiency.
 
This work was concerned solely with the assessment of the aerodynamic performance of rigid flapping wings when considering different shapes. Implementing a full aeroelastic framework that couples a nonlinear shell model and UVLM to test the obtained optimal shapes and check their superiority over the baseline configuration is the topic of our current research effort.

\section*{Acknowledgment}
The authors thank Prof. Svanberg who kindly provided the optimization package GCMMA.

\small
\bibliographystyle{elsarticle-num}
\bibliography{Ref}

\clearpage
%

\end{document}